\documentclass[aps,floats,twocolumn,superscriptaddress,floatfix,nofootinbib]{revtex4}
\usepackage{amsfonts,amssymb,amsmath,bm}
\usepackage{graphics,epsfig}

\usepackage{here}
 \newcommand{\ue}{\mathrm{e}}
\newcommand{\ui}{\mathrm{i}\,} \newcommand{\tJ}{\tilde J}
\newcommand{\tz}{\tilde{\bm z}} \newcommand{\tpartial}{\tilde\partial}
\newcommand{\tnabla}{\tilde\nabla}

\def\ds{\displaystyle}
  \def\rset{{\mathbb R}}
\def\cset{{\mathbb C}} \def\un{\hbox{{1\kern -0.25em\raise
0.4ex\hbox{{\scriptsize $|$}}}}} 
\def\nset{\hbox{{I\kern -0.18em N}}} 
  \def\v{{\bm v}} 
\def\x{{\bm x}} \def\y{{\bm y}} \def\z{{\bm z}} 
\def\k{{\bm k}} \def\h{{\bm h}}   
\def\n{{\bm n}} 
\def\twop{\boldsymbol{\mathfrak{p}}}
\def\twoq{\boldsymbol{\mathfrak{q}}} \def\pf{\mathfrak{p}}
\def\qf{\mathfrak{q}}
\def\twor{\boldsymbol{\mathfrak{r}}} \def\rf{\mathfrak{r}}

\begin{document}

\newif\iffigs \figstrue
%\figsfalse
\iffigs \fi
\def\drawing #1 #2 #3 {
\begin{center}
\setlength{\unitlength}{1mm}
\begin{picture}(#1,#2)(0,0)
\put(0,0){\framebox(#1,#2){#3}}
\end{picture}
\end{center} }

\title{Nature of complex singularities for the 2D Euler equation}
\author{W. Pauls}
\affiliation{CNRS UMR 6202, Observatoire de la C\^ote d'Azur, BP 4229, 06304 Nice Cedex 4, France}
\affiliation{Fakult\"at f\"ur Physik, Universit\"at Bielefeld,
Universit\"atsstra{\ss}e 25, 33615 Bielefeld, Germany}
\author{T. Matsumoto} \affiliation{Dep. Physics, Kyoto University,
Katashirakawa Oiwakecho Sakyo-ku, Kyoto 606-8502, Japan}
\affiliation{CNRS UMR 6202, Observatoire de la C\^ote d'Azur, BP 4229, 06304 Nice Cedex 4, France}
\author{U. Frisch}
\affiliation{CNRS UMR 6202, Observatoire de la C\^ote d'Azur, BP 4229,
06304 Nice Cedex 4, France}
\author{J. Bec} \affiliation{CNRS UMR 6202, Observatoire de la C\^ote
d'Azur, BP 4229, 06304 Nice Cedex 4, France} 
\draft{{\it Physica} D \textbf{219} (2006) 40--59} 
\date{\today}

\begin{abstract}
A detailed study of complex-space singularities of the
two-dimensional incompressible Euler equation 
is performed in the  short-time asymptotic r\'egime when such singularities are
very
far from the real domain; this allows an exact
recursive determination of arbitrarily many spatial Fourier
coefficients.  Using  high-precision arithmetic we find
that the Fourier coefficients of the stream function are given over
more than two decades of wavenumbers by $\hat F(\k) = C(\theta)
k^{-\alpha} \ue ^ {-k \delta(\theta)}$, where $\k = k(\cos \theta,\,
\sin \theta)$.  The prefactor exponent $\alpha$, typically between
$5/2$ and $8/3$, is determined with an accuracy better than $0.01$. It
depends on the initial condition but not on $\theta$. The vorticity
diverges as $s^{-\beta}$, where $\alpha+\beta= 7/2$ and $s$ is the
distance to the (complex) singular manifold. This new type of
non-universal singularity is permitted by the strong reduction of
nonlinearity (depletion) which is associated to  incompressibility. 
Spectral calculations show that the scaling reported
above persists well beyond the time of validity of the short-time
asymptotics.
A simple model in which the vorticity is treated as a
passive scalar is shown analytically to have universal singularities
with exponent $\alpha =5/2$.

\end{abstract}
\vspace*{5mm}
%]
\maketitle

{\em \noindent Und es wallet und siedet und brauset und zischt,\\ Wie
wenn Wasser mit Feuer sich mengt,\\ Bis zum Himmel spritzet der
dampfende Gischt,\\ Und Flut auf Flut sich ohn' Ende dr\"angt...} \\[1ex]
%~\\
\noindent Friedrich von Schiller, from {\em Der Taucher} \cite{poem}

%%%%%%%%%%%%%%%%%%%%%% INTRODUCTION %%%%%%%%%%%%%%%%%%%%%%%%%%%%%%%%%%

\section{Introduction}
\label{s:introduction}

A quarter of a millenium has elapsed since Euler published for
the first time what is now known as the Euler equations of
hydrodynamics \cite{Euler}. There has not been much celebration but
this may just reflect our embarrassment at not having made enough
progress. Actually, Leonhard Euler warned us. At the end of his 1755
paper he wrote: ``However all that the Theory of fluids holds, is
contained in the two equations above, so that in the pursuit of the
research we are not lacking the principles of Mechanics, but solely
the Analysis, which is not yet cultivated enough for this design:
hence we see clearly, which discoveries are left for us to make in
this Science, before we can attain a more perfect Theory of the motion
of fluids''.\footnote{In French: Cependant tout ce que la Th\'eorie
des fluides renferme, est contenu dans les deux \'equations
rapport\'ees cy-dessus, de sorte que ce ne sont pas les principes de
M\'echanique qui nous manquent dans la poursuite de ces recherches,
mais uniquement l'Analyse, qui n'est pas encore ass\'es cultiv\'ee,
pour ce dessein: et partant on voit clairement, quelles d\'ecouvertes
nous restent encore \`a faire dans cette Science, avant que nous
puissions arriver \`a une Th\'eorie plus parfaite du mouvement des
fluides.} (A paper in Latin \textit{Principia motus
fluidorum}, published a few years after the paper in French, contains
the basic equations and was  already presented under a 
different title to the Berlin Academy in 1752.)

Euler considered both the compressible and incompressible cases. Here
we are concerned only with the latter which is particularly difficult
in view of the global nature of the incompressibility constraint.  One
of the most important open questions concerning the ``analysis'' of
the Euler equations is the well-posedness: does initially smooth 3D
flow, which is known to remain smooth for short times, eventually
``blow up'', that is become singular in a finite time (see, e.g.\
Refs.~\cite{MB,blue})?  In two dimensions it has been known since the
thirties that flow in a bounded domain, initially sufficiently smooth,
never blows up \cite{Hoelder,Wolibner}. It was also shown that if such
a 2D flow is initially analytic it will stay so forever 
\cite{BBZ,Benachoura,Benachourb}. However, in the course of time, such
flow can develop very fine scales and there is a large discrepancy
between the analytic estimation of how the smallest scale decreases in
time (a double exponential) and what is found in numerical simulations
(a simple exponential; see, e.g., Ref.~\cite{SSF}). 

The likely cause
of the discrepancy is \textit{depletion}, the phenomenon by which
high-Reynolds number or inviscid incompressible flow tends to organize
itself into structures having vastly reduced nonlinearities (see,
e.g., Ref.~\cite{cup}). Depletion, which is still very poorly
understood, may hold the key for understanding why 3D high-Reynolds number flow
seems never to blow up, at least in simulations.\footnote {For the
case of 3D inviscid Euler flow there is no truly conclusive evidence
in favor of blow up \cite{blue,CB}. Furthermore, if the flow is
initially analytic, any real singularity will have to be preceded by
complex-space singularities \cite{Benachoura,Benachourb}.}  
In this paper we shall focus on the two-dimensional case.

There are well-known 2D examples of depletion, such as flows which
depend only on one Cartesian coordinate or on the radial polar
coordinate.  Such flows are however steady and thus globally depleted,
with no dynamics. In this paper we shall be interested in 
2D flow with an initial stream function which is a real trigonometric 
polynomial in the space variables, of the sort already considered in
Refs.~\cite{blue,FDR}. These are the 2D counterparts of well-known 3D
flows such as Taylor--Green and Kida--Pelz \cite{TG,BMONMF,Kida,Pelz,CB} which
have been used for (so far inconclusive) investigation of finite-time
blow up. Our 2D flows have generally non-trivial dynamics and
display locally very strong depletion.

Trigonometric polynomials are instances of entire functions, that is,
functions which are analytic in the whole complex domain. 
The only singularities of such functions
are at complex infinity. The solution of the Euler equations
at times $t>0$ sufficiently small can then be extended analytically into
the complex domain \cite{BBZ,Benachoura,Benachourb}. There is
strong numerical evidence in  2D and also in 3D that such flow
does not stay entire and develops singularities at certain complex locations
for any $t>0$ \cite{SSF,BMONMF,blue,FDR}. Complex singularities
are usually detected through the Fourier transforms
of the solution: roughly, there is an exponential tail related to the
distance of the nearest singularity from the real domain, accompanied by an
algebraic prefactor related to the nature (also called type or
structure).of the singularities.

Little is known about the nature of complex singularities of the Euler 
equations.  In 
Refs.~\cite{blue} and \cite{FDR} it is shown numerically for the 2D case
with the initial stream function $\cos x_1+\cos 2x_2$ that 
the complex singularities lie on a smooth manifold and that the
vorticity becomes infinite when approaching the singular
manifold; there is however considerable uncertainty as to the scaling
law of this divergence. In Ref.~\cite{TS93} the motion of preexisting
complex-space singularities is studied analytically but their nature
is kept quite arbitrary. In Ref.~\cite{caflisch} traveling-wave solutions
with a pure imaginary velocity are studied for 3D axisymmetrical flow
with swirl; using an ultra-high precision\footnote{that is, higher
than double precision.} numerical method, the singularities in the
complexified axial variable are mapped out as a function of the (real)
radial variable and found to lie on a smooth curve; the nature of the
singularities is obtained using a ``sliding fit'' method. In
Ref.~\cite{shelley}, for the vortex sheet problem with an initially
analytic interface, the nature of complex singularities of the
interface is obtained using an ultra-high precision method and a
``pointwise fit''. The sliding fit and the pointwise fit are very
closely related to the method we use in Section~\ref{ss:captalpref} and
we shall come back to this matter. Since the work of Krasny \cite{krasny},
it appears that ultra-high precision is a prerequisite for obtaining
numerical information on the nature of singularities, particularly when they are
in the complex domain.

From a \textit{theoretical} point of view, for many nonlinear
equations of mathematical physics a very successful tool in studying
the nature of singularities has been dominant balance and its refined
versions such as Painlev\'e analysis \cite{WTC}.
Dominant balance analysis typically gives \textit{universal}
singularities, that
is singularities whose positions may depend on the initial conditions
but their nature does not. 
The simplest instance is the 1D viscous Burgers
equation whose complex-space singularities are simple poles, obtainable
by balancing the nonlinear term against the viscous one. 
For the $d$-dimensional incompressible
Euler equations, attempts to use dominant balance fail because of the
particular structure of the nonlinearity: if we assume that the
solution becomes singular on a complex manifold of dimension $d-1$,
the nonlinearity vanishes to leading order. This is just a consequence
of the simplest form of depletion, the vanishing of nonlinearity for
solutions which depend on a single spatial coordinate.  The nature of
singularities cannot be obtained by a dominant balance argument;
actually, as we shall see, complex singularities of the 2D Euler equation
display a very unusual \textit{non-universality}.

In this paper we will mainly discuss the \textit{short-time asymptotic
r\'egime} presented in Ref.~\cite{blue} and extensively used in
Ref.~\cite{FDR} which gives us the most accurate information on the
nature of complex singularities.\footnote{Henceforth, Ref.~\cite{FDR}
will be cited as MBF.}  After briefly introducing it in
Section~\ref{s:pshydr} we will show that this r\'egime can be
reformulated as a ``pseudo-hydrodynamic'' Euler problem, in which all
the action including the singularities takes place in a plane extending in the
pure imaginary directions, but our usual
hydrodynamic intuition is still applicable.  The short-time
asymptotics allows us to obtain recurrence relations for spatial
Fourier components involving only wavectors $\k =(k_1,\,k_2)$ with
$k_1 \ge 0$ and $k_2\ge 0$, a feature which is also present in the
Moore approximation for vortex sheets \cite{moore} and its
generalization to smooth flow \cite{caflisch}; as a consequence Fourier
components can be calculated in ultra-high precision 
without any truncation error.
%\footnote{In
%Ref.~\cite{caflisch}  up to  128 decimal figures are used; here we use
%up to 100 decimal figures.} 
In Section~\ref{s:fourier} we present the numerical evidence
for simple scaling laws associated to complex singularities and
determine the nature of the singularities with high precision. 
Analyzing short-time asymptotics for
different initial conditions, we find that the singularities 
are non-universal.  In Section~\ref{s:geometry} we
describe the global and local geometry of the pseudo-hydrodynamic
flow, including depletion of nonlinearity which is especially strong
near the singularities.  

Sections~\ref{s:fourier} and \ref{s:geometry}
both involve a mixture of numerical results and of theoretical
arguments, some heuristic, some more rigorous. We must stress that at
the moment we do understand various features of the solution, in
particular why the scaling exponent for singularities does not depend
on the direction, but we failed so far to reproduce by theory the
non-universal scaling exponents observed for the
singularities. Nevertheless by moving to yet another level of
toy-modeling (the equivalent for our problem of considering the
vorticity as a passive scalar in a prescribed velocity field), we can
determine the nature of the corresponding complex singularities using
dynamical systems tools (Section~\ref{s:linearized}). The nature of
these ``advection'' singularities is however universal and therefore
does not reproduce an essential feature of the nonlinear Euler flow.
Finally, conclusions, open problems and a tentative road map  for
future research on blow up are presented in
Section~\ref{s:conclusion}. To make the present paper reasonably
self-contained we shall occasionally re-derive results already found in
Ref.~\cite{blue} and MBF.
%\vfill \strut
%%%%%%%%%%%%%%%%%%%%%% SECTION 1 %%%%%%%%%%%%%%%%%%%%%%%%%%%%%%%%%%

\section{Short-time asymptotics and pseudo-hydrodynamics}
\label{s:pshydr}

We are interested in the short-time asymptotics for the 2D Euler
equation, written in terms of the stream function
\begin{equation} \label{e:euler}
\partial _t \nabla ^2 \Psi (\x,t) - J (\Psi , \nabla ^2 \Psi )=0\; ,
\end{equation}
where $\x = (x_1,x_2)$ and $J(f,g)\equiv \partial_1f\partial_2g
-\partial_1g\partial_2f$. The initial condition $\Psi_0(\x)\equiv
\Psi(\x,0)$ is a real $2\pi$-periodic trigonometric polynomial of the
form $\Psi_0(\x) = \sum_{\k} \hat F^{(0)}(\k)\, \ue ^{\ui\k\cdot\x}$,
where the sum has only a finite number of terms.  Here
$\k=(k_1,k_2)$, where $k_1$ and $k_2$ are signed integers.  The
short-time asymptotics is simplest when the initial condition has only
two orthogonal Fourier modes, as in Refs.~\cite{blue,FDR} where the
assumed initial condition is
\begin{equation}
\Psi_0(\x) =\cos x_1 +\cos 2x_2\;.
\label{proto}
\end{equation}
In what follows we shall call this initial condition Standard
Orthogonal Case (SOC). One of our present goals is
to investigate to what extent complex singularities are or are not
universal, we are thus naturally led to considering more general cases,
having, e.g., more than two modes in the initial conditions. In
Appendix~\ref{a:multimode} it will be shown that the short-time
asymptotic r\'egime for the multimode case
can be reduced to a set of two-mode initial
conditions. We may thus without loss of
generality limit ourselves to two-mode initial conditions of the form
\begin{equation}
\Psi _0 (\x) = h_1 \, \ue ^{\ui \twop \cdot \x } + h_2 \, \ue ^{\ui \twoq
\cdot \x } + {\rm c.c.}\;.
\label{two_mode}
\end{equation}
Here c.c. stands for ``complex conjugate'', $\twop = (\frak{p} _1 ,
\frak{p} _2 ) $ and $\boldsymbol{\mathfrak{q} } = (\frak{q} _1 ,
\frak{q} _2 ) $ are two vectors with signed integer components. 
Furthermore, we assume
that $\twop$ and $\twoq$ are not parallel and do not have the same
modulus since otherwise the two-mode initial condition is a
time-independent solution of the Euler equation. By performing if
needed
a suitable translation, we can then assume that $h_1$ and $h_2$ are
positive. Finally, since our
goal here is primarily to demonstrate non-universality of the nature
of the singularities with respect to the initial conditions, we shall
not strive for the greatest generality and limit ourselves to \textit{basic
modes} with  non-negative components such that $\mathfrak{p} _1
\frak{q} _2 - \frak{q} _1 \frak{p} _2 >0$.

Eq.~\eqref{e:euler} has a solution in the form of
a Taylor series in the time variable
\begin{equation}
  \Psi(\x,t) = \sum_{n\ge0} \Psi_n(\x)\, t^n\;,
  \label{shorttimeexpansion}
\end{equation}
where $\Psi_0$ is the initial condition and the $\Psi_n(\x)$'s for
$n\ge 1$ are easily shown to satisfy the  recursion relations:
\begin{equation}
  \nabla^2\Psi_{n+1} = \frac{1}{n+1} \sum_{m+p=n} J(\Psi_m,
  \nabla^2\Psi_p)\;.
  \label{recurrencepsin}
\end{equation}
For the two-mode initial condition \eqref{proto} all the $\Psi_n(\x)$
are trigonometric polynomials that can be continued analytically to
complex locations $\z =\x +\ui \y$. Since the initial condition has
its singularities at infinity, we expect, by continuity, that at short
times  the singularities will have large imaginary parts $\vert y_1
\vert $ and $\vert y_2 \vert $. Let us now suppose that $y_1 \to
+\infty$ and $y_2 \to +\infty$ in such a way that their ratio $y_2 / y_1$
stays finite but arbitrary. Obviously, the four vectors $(\frak{p}
_1 , \frak{p} _2 )$, $(-\frak{p} _1 , -\frak{p} _2)$, $(\frak{q} _1 ,
\frak{q} _2 )$ and $(-\frak{q} _1 , - \frak{q} _2 )$ divide the $\k
$-space into four angular sectors so that, e.g.\ in the first angular
sector $ \frak{p} _2 /\frak{p} _1 \leq y_2 / y_1 \leq \frak{q} _2 /
\frak{q} _1 $. Then, for $\z $ such that $\y $ lies in the first
angular sector,  to leading order any additional
factor $t$ in the expansion \eqref{shorttimeexpansion} is
accompanied by either a factor $\ue ^{-i\twop \cdot \z }$ or a factor
$\ue ^{-i\twoq \cdot \z }$, thus giving amplitude factors $t \ue ^{
\twop \cdot \y } $ and $t \ue ^{ \twoq \cdot \y } $, respectively.
When $t \rightarrow 0 $ and $|\y| \rightarrow \infty $ these factors
remain finite, provided $ \twop \cdot \y $ and $\twoq \cdot \y $ are
shifted by $\ln t $. This suggests that the short-time asymptotics is
obtained by the \textit{similarity ansatz}
\begin{eqnarray}
&&\Psi(\z,t) = (1/t) F(\tz)\;,
\label{simansatz}\\
&&\tz = (\tilde z_1,\, \tilde z_2) \equiv (z_1+\ui \lambda _1 \ln t,\,
z_2+ \ui \lambda _2 \ln t)\; ,
\label{deftz}
\end{eqnarray}
where $\lambda _1 $ and $\lambda _2 $ are determined by
\begin{equation}
\lambda _1 = \frac{\frak{p} _2 - \frak{q} _2 }{\frak{q} _1 \frak{p} _2
  - \frak{q} _2 \frak{p} _1 } \quad \mathrm{and} \quad \lambda _2 =
\frac{\frak{q} _1 - \frak{p} _1 }{\frak{q} _1 \frak{p} _2 - \frak{q}
  _2 \frak{p} _1 } \;.
\label{scaling_factors}
\end{equation}  
Substitution in \eqref{e:euler} gives the \textit{similarity equation}
\begin{equation}
\tnabla ^2 (-1+\ui \lambda _1 \tpartial_{z_1}+ \ui \lambda _2
\tpartial_{z_2}) F = \tJ(F, \tnabla ^2 F)\;,
\label{asympteuler}
\end{equation}
where the  tilde means that the partial derivatives are
taken with respect to the new variables. The initial condition
\eqref{proto} becomes an asymptotic boundary condition
\begin{equation}
F(\tz) \simeq h_1 \ue ^{-\ui \twop \cdot \tilde{\z } } + h_2 \ue
^{-\ui \twoq \cdot \tilde{\z } } , \quad \tilde y_1\to
-\infty,\,\,\,\,\tilde y_2\to -\infty\;.
\label{asymbound}
\end{equation}
In \eqref{asympteuler} the second and third terms on the l.h.s.\ can
be viewed as stemming from the advection by a pure imaginary constant
``drift velocity''.  This is because we are following the singularities coming
``down'' from complex infinity. It is important to observe that
\eqref{asympteuler} is an exact consequence of the Euler equation. 
The only place where an
approximation is made is in the boundary condition \eqref{asymbound}
where harmonics containing e.g.\ $\ue ^{+\ui \twop \cdot \tilde{\z }
}$ and $\ue ^{+\ui \twoq \cdot \tilde{\z } } $ are discarded because
such terms are exponentially subdominant at short times.

In what follows we shall generally limit ourselves to SOC, giving
occasionally an indication of what is valid for more general two-mode
cases. The general case can easily be handled but we wish to avoid
burdening the reader with unnecessarily complicated statements and
equations.
 
The function $F(\tz)$, which is $2\pi$-periodic in $\tilde x_1$ and
$\pi$-periodic in $\tilde x_2$, is analytic in the product of the
half-spaces $\tilde y_1 \le 0$ and $\tilde y_2 \le 0$ and thus its
spatial Fourier series has only harmonics of the form $\ue ^{-\ui
(k_1\tz_1+k_2\tz_2)}$ with $k_1\ge 0$ and $k_2\ge 0$.  Its Fourier
series is here written as\footnote{In MBF $k_1$ and $k_2$ were defined
with the opposite sign.}
\begin{equation} \label{e:fourier}
F(\tz_1 ,\tz_2 ) = \sum_{k_1 = 0}^{\infty } \sum_{k_2 = 0}^{\infty }
(-1)^{k_1} \hat{F} (k_1 , k_2 ) \ue ^{-\ui k_1 \tz_1 } \ue ^{-\ui k_2
\tz_2 }\;.
\end{equation}
The reason for the presence of the factor $(-1)^{k_1}$ will become
clear shortly. The Fourier coefficients $\hat F(\k) \equiv \hat
F(k_1,k_2)$ can be calculated recursively from the relation given in
MBF which follows from \eqref{asympteuler}
\begin{eqnarray} \label{e:difference}
&&\hat{F} (k_1 , k_2 ) = - \frac{1}{k_1 + k_2/2 - 1}\, \frac{1}{\vert
{\bf k} \vert ^2 } \times \\ &&\sum_{p_1 = 0}^{k_1 } \sum_{p_2 =
0}^{k_2 } ({\bf p} \wedge {\bf k}) \vert {\bf k} - {\bf p} \vert ^2
\hat{F} (p_1 , p_2 ) \hat{F} (k_1 - p_1 , k_2 - p_2 )\;.\nonumber
\end{eqnarray}
Here, ${\bf p} \wedge {\bf k} \equiv p_1 k_2 - p_2 k_1 $. Because
there are no Fourier harmonics with negative $k_1$ or $k_2$, the 
convolutions in \eqref{e:difference} only involve positive
arguments. This feature, which allows truncation-free determination
of Fourier coefficients,  is also present in the Moore approximation
for the vortex sheet problem and in its generalization to axisymmetrical flow
\cite{moore,caflisch}. The
initialization of the recursion relations requires the knowledge of the
coefficients along the ``edges'', that is the half-lines $k_1=0$ and
$k_2 =0$.  In the present case $\hat{F} (0,2) = 1/2 $ and 
$\hat{F} (1,0)= - 1/2 $ while  all the other edge harmonics are 
zero.\footnote{In the
general case of two basic modes $\twop $ and $\twoq $, the main change
with respect to SOC is the replacement in \eqref{e:difference} of the
denominator $k_1+k_2/2 -1$ by $\lambda _1 k_1 + \lambda _2 k_2 - 1$
where $\lambda _1$ and $\lambda _2$ are defined in
\eqref{scaling_factors}.} It has been shown in MBF that, with the
choice made above in \eqref{e:fourier}, the coefficient $\hat{F}
(1,0)$ is the only one that is negative. All the other ones are
non-negative. This result has so far only been established by (very
solid) numerical computations and holds for all the two-mode initial
conditions studied. As we shall see, this has important
consequences for the geometry of singularities.

We shall now show that \eqref{asympteuler} can be reformulated as the
steady solution of a \textit{pseudo-hydrodynamic} problem in a
suitable imaginary plane.  Since we are working with analytic
functions, we can replace the complex partial derivatives
$\tpartial_{z_1}$ and $\tpartial_{z_2}$ by $-\ui \tpartial_{y_1}$ and
$-\ui \tpartial_{y_2}$, holding the $x$-coordinates fixed. In terms of
such $y$-derivatives \eqref{asympteuler} becomes an equation with real
coefficients. If we furthermore choose $x_1$ and $x_2$ such that the
boundary condition \eqref{asymbound} becomes real then the solution
``above such points'' $F$ is also real.  This happens for $x_1 =0,\,
\pi$ and for $x_2 =0,\, \pi/2,\,\pi,\, 3\pi/2$. The positivity of all
but one of the Fourier coefficients defined in \eqref{e:fourier} with
the factor $(-1)^{k_1}$, amounts to stating that, after moving the
origin to $(\pi,\,0)$, all but one of the usual Fourier coefficients
of $F$ are positive. As we shall see in Section~\ref{s:geometry}, this
gives us the possibility of analyzing the (short-time) complex
singularities by focusing solely on the $y$-plane above
$(\pi,\,0)$. This point turns out also to be a center of symmetry for
the Euler flow with the initial condition \eqref{proto}, but it is not
clear whether this matters.\footnote{Note that streamlines have a
hyperbolic structure near $(\pi,\,0)$, but an elliptic structure near
$(0,\, 0)$ which is also a center of symmetry.}  Henceforth we shall
consider the $y$-plane above $(\pi,\,0)$.

We define a pseudo-stream function in terms of the $y$-coordinates
(from now on we drop the tilde on the $y$ variables for notational
simplicity)
\begin{eqnarray}
\psi(\y) &=& \psi(y_1,y_2) \equiv \frac{1}{2} y_1-y_2 + F(\pi + \ui
y_1, \ui y_2)\;,\label{defpsi}\\ &=&\frac{1}{2} y_1-y_2 +\sum_{\k}\hat
F(\k)\,\ue ^{\k\cdot\y}\;,
\label{psifourierlaplace}
\end{eqnarray}
where the two linear terms on the r.h.s.\ have been introduced to avoid
having an additional advection term. Note that because of these terms,
$\psi$ is not the continuation to complex coordinates of a function
periodic in $x_1$ and $x_2$.

It is now elementary to check that
\eqref{asympteuler}-\eqref{asymbound} 
are equivalent
to taking the steady-state ($\tau$-independent) solution of the
pseudo-hydrodynamic equation
\begin{equation}
\partial_\tau \nabla ^2 \psi -J(\psi, \nabla ^2 \psi) = - \nabla ^2
\psi\;,
\label{pseudopsi}
\end{equation} 
with the asymptotic boundary condition (for $y_1,\, y_2 \to -\infty$)
\begin{equation}
\psi(y_1,y_2) -{1\over 2} y_1 +y_2 \simeq  -{1\over 2}\ue ^{y_1} +
{1\over 2} \ue ^{2y_2}\;.\label{boundarypseudopsi}
\end{equation}
Here, in order to bring out familiar hydrodynamic notation, we have
introduced a pseudo-time variable $\tau$.\footnote{If we allow
the function $F$ and thus $\psi$ to also depend on $t$ and set $\tau = \ln (1/t)$, we obtain
precisely \eqref{pseudopsi}.} We are using $\nabla
=(\partial_1,\,\partial_2)$ for $\nabla_y$ and the Jacobian $J$ has
its usual definition in terms of $y$-derivatives.  We now introduce a
pseudo-velocity and a pseudo-vorticity by the usual
definitions:\footnote{The true velocity is actually pure imaginary in
the $y$-plane and the true vorticity is $-\omega$.}
\begin{eqnarray}
\v &=&(v_1,\,v_2) \equiv (\partial_2 \psi,\, -\partial_1 \psi)
\label{defpseudov}\\
&=& -(1,\, 1/2) + \sum_{\k}\,(k_2,\, -k_1)\,\hat F(\k)\,\ue
^{\k\cdot\y}\;,
\label{vfourierlaplace}\\
\omega &\equiv & -\nabla ^2 \psi\;,
\label{defpseudoomega}\\
&=& \sum_{\k} -k^2\,\hat F(\k)\,\ue ^{\k\cdot\y}\;,
\label{omegafourierlaplace}
\end{eqnarray}
in terms of which \eqref{pseudopsi} reads
\begin{equation}
\left[ \partial_\tau \omega\right] +
\v \cdot\nabla \omega +\omega=0\;,
\label{pseudoomega}
\end{equation}
with the boundary conditions (for $y_1,\, y_2 \to -\infty$)
\begin{equation}
\v \simeq \left(-1,\, -{1\over2}\right),\quad \omega \simeq {1\over2}\ue ^{y_1}
-2\ue ^{2y_2}\;.
\label{boundvomega}
\end{equation}
For other initial conditions, only the boundary condition
\eqref{boundvomega} must be modified.
The $\tau$-derivative term has been put within square brackets since
we are only interested in the steady-state solution.  Note that the
pseudo-hydrodynamic formulation in the $y$-plane is that of a
quasi-two-dimensional flow in a 3D container with bottom friction
producing a Rayleigh drag. In this formulation $\tau \to +\infty$
as we approach the initial instant. An alternative interpretation is to define
$\tau$ as $\ln t$, to avoid reversing the course of time, and then
to change the signs of $\v$ and of $\omega$ and replace  the Rayleigh drag by
an instability.

In the pseudo-hydrodynamic  formulation it  is now obvious that the problem
is invariant under an arbitrary translation $\h = (h_1,\,h_2)$ in $y$-space. 
By \eqref{psifourierlaplace}, such a translation amounts to a factor 
$\ue ^{\k\cdot\h}$ on the Fourier coefficients $\hat F(\k)$. It follows, as
noted in MBF,  that the set of initial conditions $\Psi_0(\x) =
\ue ^{h_1} \cos x_1 + \ue ^{2h_2} \cos 2x_2$ is equivalent to SOC as
long as $\h$ is within the analyticity domain. Similarly, a translation
in $k$-space with integer components  $(n_1,\, n_2)$ is equivalent to
multiplying $F(\pi+ \ui y_1,\, \ui y_2)$ by the exponential factor 
$\ue ^{n_1y_1+n_2y_2}$ in $y$-space. The exponential being an entire function, 
this changes neither the positions nor 
the nature of the singularities at finite distance.

\section{Numerical investigation of scaling laws in Fourier space}
\label{s:fourier}

We shall show in this section that the solution of the Euler equation
in the short-time asymptotic r\'egime defined in the previous section,
has remarkably clean scaling properties in Fourier space. By this we
mean that the wavenumber dependence of the Fourier coefficients is
represented as a decreasing exponential multiplied by an algebraic
prefactor whose exponent can be measured very accurately. Such a
functional form is not surprising. In fact the exponential is the
signature of the location of a singularity while the prefactor
encodes the nature of the singularity. For one-dimensional
analytical functions with isolated singularities in the complex space
this is well known: a singularity at $z_\star$ of the form
$(z-z_\star) ^{\rho}$ has a signature in the modulus of the Fourier 
transform at high
wavenumbers $k$  of the form $C|k|^{ -\rho -1} \ue ^{-\delta |k|}$,
where $\delta$ is the the distance of $z_\star$ to the real axis (see,
e.g., Ref.~\cite{CKP}). Such asymptotic results have been extended in
the nineties to the Fourier transforms of periodic 
analytical functions of several complex variables
when the wavevector $\k$ tends to infinity
with a fixed rational slope $\tan \theta = k_2/k_1 =p/q$, where $p$
and $q$ are relative prime integers\cite{Tsikh,Orlova,Orlovb}.

When the Fourier coefficients are obtained numerically, there is a
maximum wavenumber $k_{\rm max}$. Unless it is taken very large,
there will be very few points on the line of slope $p/q$ as soon as
$q$ is not a very small integer. But a large value of $k_{\rm max}$
entails extremely small Fourier coefficients because of the
exponential decrease with the wavenumber. Thus, as stressed in MBF,
very high precision may be
needed to avoid swamping by the rounding errors.  Truncation errors
are not an issue in the short-time asymptotic r\'egime since the Fourier
coefficients can be calculated from \eqref{e:difference} with arbitrary
accuracy.  

The data obtained for the SOC initial condition in MBF had
wavenumbers $k \equiv |\k|$ up to 1000 or 2000, depending on the
direction and were calculated with 35-digit accuracy.\footnote{In MBF
it was stated that, when using only double-precision (15-digit)
accuracy, unacceptably large errors are obtained beyond wavenumber
800. Actually, as pointed out by P.~Zimmermann (private communication),
the double-precision calculation can be modified  in  such a way that,
up to wavenumber 1000, the relative error on Fourier modes does not
exceed $10^{-5}$.} Most of the results  presented here are based on
the 35-digit calculation.  Additional calculations
are also presented here with various initial conditions, with up to
100-digit precision and wavenumbers which can reach 4000 in particular
directions. We note that the MPFUN90 package for high-precision
calculation used in MBF, here and in Ref.~\cite{caflisch} 
makes use of fast Fourier transform
techniques. Thus the CPU time per multiplication, as a function 
of the number of digits $N$, is proportional to $N \log N$
\cite{mpfun}.

\begin{figure}[t]
 \iffigs 
 \centerline{%
 \includegraphics[scale=0.7]{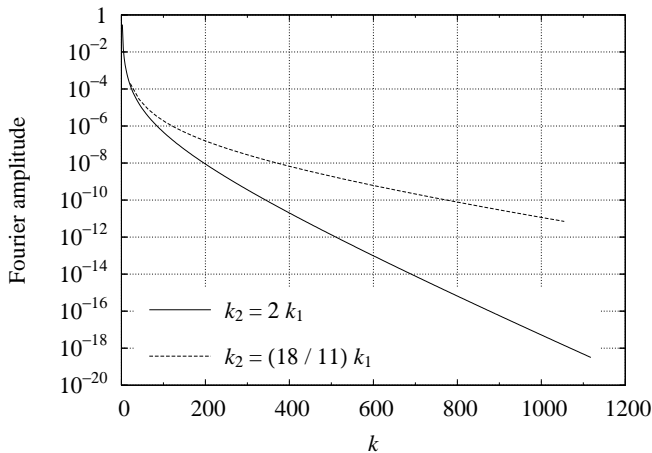}%
 }
 \else\drawing 65 10 {soc lin-log 2/1}
 \fi 
 \caption{\label{f:socslope2a}Fourier coefficients of the stream 
function $F$ along  two lines of different slopes as a function of $k \equiv |\bm k|$ in lin-log coordinates.} 
\end{figure}
\begin{figure}[t]
 \iffigs  
  \centerline{%
 \includegraphics[scale=0.7]{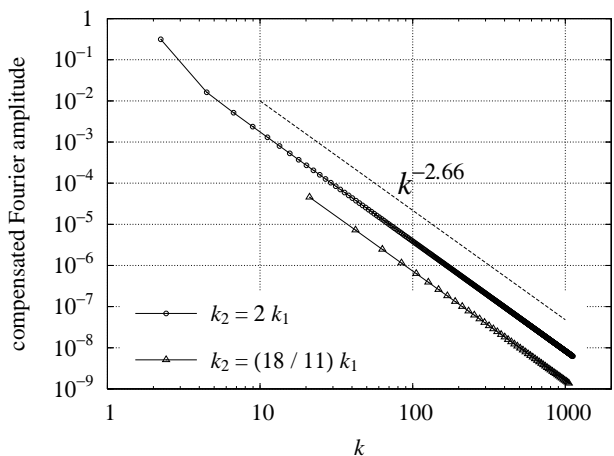}%
 }
 \else\drawing 65 10 {soc compen log-log 2/1}
 \fi  
 \caption{\label{f:socslope2b}Same as in Fig.~\ref{f:socslope2a} after
division by $\exp(-\delta k)$ (compensated Fourier coefficients) in
log-log coordinates. Most of  the points are in the
asymptotic  power-law r\'egime, at least visually.} 
\end{figure}

We now show that it is quite easy in principle
to observe scaling by  analyzing the behavior
of the Fourier coefficients in directions of rational slope. 
Figs.~\ref{f:socslope2a} and \ref{f:socslope2b} give two examples of  the
analysis of Fourier coefficients along straight lines through the 
origin\footnote{All the  lattice lines of a given rational slope have the same
high-$k$ asymptotics, due to the observation made at the end
of Section~\ref{s:pshydr}.} in a direction of rational slope,
using the data from MBF for the Fourier coefficients of the stream
function with SOC initial conditions. 
The first  case has  $k_2 / k_1 =2$, the direction with the
largest number of grid points having non-vanishing Fourier coefficients.
The second case has $k_2/k_1 =18/11$, the direction with the slowest
decrease of the Fourier coefficients.
Fig.~\ref{f:socslope2a}, which shows the Fourier coefficients in lin-log
coordinates,reveals an exponential tail $\propto e ^{-\delta
k}$; a least square fit gives $\delta = 0.021$  for the first case and
$\delta = 0.0065$ for the second case.\footnote{Why the minimum value is so 
small  is
a matter we shall come back to in Section~\ref{s:linearized}.}
In Fig.~\ref{f:socslope2b} we show
the ``compensated'' Fourier coefficients obtained by dividing by the
exponential term; the result is then represented in log-log
coordinates in order to look for an algebraic prefactor $\propto
k^{-\alpha}$. 
The quality of the scaling obtained is impressive: over most of the range
we cannot on a log-log plot
visually distinguish the prefactor  
from a power law  with exponent $\alpha = 8/3$. 
As we shall see, the exponent does not depend on the direction chosen.

\subsection{Technique for capturing algebraic prefactors}
\label{ss:captalpref}

Determining the scaling properties as done above by use of least
square fits, compensating exponentials and log-log plots is not
optimally adapted for delicate issues such as studying the dependence of
the prefactor exponent on the direction of the wavector or on the
initial conditions. As pointed out by Shelley \cite{shelley}, it is
better to remove some of the subjective biases present in a least
square fit (such as choosing the range in $k$). We shall make use of
his method of point-wise fit (also used in Ref.~\cite{caflisch}, where
it is called a sliding fit), followed by an extrapolation step as now
explained.
 
In $\k$-space, a direction of rational positive slope is characterized
 by $k_2 / k_1 = \tan \theta =q / p$ (where the positive integers $p$
 and $q$ are taken to be relative primes). All the $\k$ vectors on the
 line of slope $p/q$ through the origin are thus of the form $\k
 =n\k_0$, where $\k_0 \equiv (p,\,q)$ and $n$ is a positive integer.
 What we have seen at the beginning of Section~\ref{s:fourier} suggests that for a
 given direction of rational slope $\tan \theta$, the Fourier
 coefficients of the stream function can be represented, for
 sufficiently large $k$, as
\begin{equation} \label{e:precise_asymptotics}
 \hat{F} \simeq C(\theta ) k^{-\alpha (\theta ) } 
\ue ^{-\delta (\theta ) k } \;.
\end{equation}
Henceforth $\alpha$, $C$ and $\delta$ will be referred to as the
prefactor exponent, the constant and the decrement, respectively. When
there is no ambiguity, the $\theta$-dependence will be omitted.
Following
Ref.~\cite{shelley}, let
us assume for a moment that \eqref{e:precise_asymptotics} 
holds exactly and let us set
$\hat F_n(\k_0) \equiv \hat F(n\k_0)$. It then follows that if we know
$\hat F_n(\k_0)$ for any three consecutive values, say $n-1$, $n$ and
$n+1$, we can determine $C$, $\alpha$ and $\delta$ by
\begin{equation} \label{e:alpha}
\alpha = \frac{\ln \left({\ds \frac{\hat F_{n-1} (\k_0 ) \hat F_{n+1}
(\k_0 ) }{\hat F^2_n (\k_0 ) } }\right)}{\ln \left({\ds \frac{n^2
}{(n-1)(n+1)} }\right)} \;,
\end{equation}
\begin{equation} \label{e:delta}
\delta = \frac{1}{\vert \k_0 \vert } \left[ \ln \left( \frac{\hat F_n
(\k_0 ) }{\hat F_{n+1} (\k_0 ) } \right) +\alpha \ln \left(\frac{n
}{n+1 } \right) \right]\;,
\end{equation}
\begin{equation} \label{e:C}
\ln C = \ln \hat F_{n} (\k_0 ) + \alpha \ln \left[(n) \vert \k_0
\vert \right] + n \vert \k_0 \vert \delta \;.
\end{equation}
The expression \eqref{e:alpha} for $\alpha$
follows immediately by noticing that in the combination $\hat
F_{n-1}\hat F_{n+1}/\hat F^2_n$ the constant $C$ and the exponential
factor both drop out. The other two expressions are readily established
by taking the logarithm of \eqref{e:precise_asymptotics}.

Of course, we have no reason to expect that
\eqref{e:precise_asymptotics} holds \textit{exactly} for arbitrary
wavenumbers. At best it will hold asymptotically at large
wavenumbers. Nevertheless we can use \eqref{e:delta}--\eqref{e:C} to
calculate a \textit{local prefactor exponent} $\alpha_{\rm loc}(k)$, which
depends on the wavenumber. Here we have chosen to use as arguments of the local
quantities the wavenumber $k=n|\k_0|$.

\begin{figure}%[H]
 \iffigs   
  \centerline{%
 \includegraphics[scale=0.7]{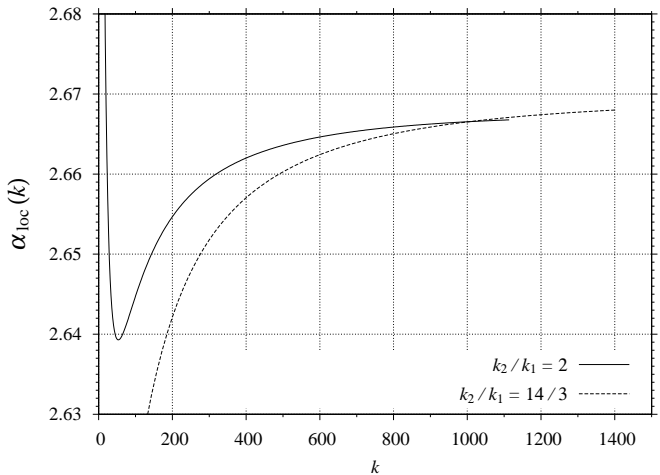}%
 }
  \else\drawing 65 10 {alpha SOC 2 examples}
 \fi 
 \caption{\label{f:typicalalphak}Local prefactor exponent
 $\alpha_{\rm loc}(k)$ vs wavenumber for two values of the slope.} 
\end{figure}
The typical behavior of $\alpha_{\rm loc}(k)$ is shown in 
Fig.~\ref{f:typicalalphak} for two directions. For large
values of $k$ the curves grow to an asymptotic value close to $8/3$.
Globally, $\alpha_{\rm loc}(k)$ is found to be non-monotonic when
$\theta <\theta_\star$ with $\tan \theta_\star$ close to 3 (but not
very sharply defined)  and monotonic above $\theta_\star$.

\begin{figure}%[H]
 \iffigs   
 \centerline{% 
\includegraphics[scale=0.7]{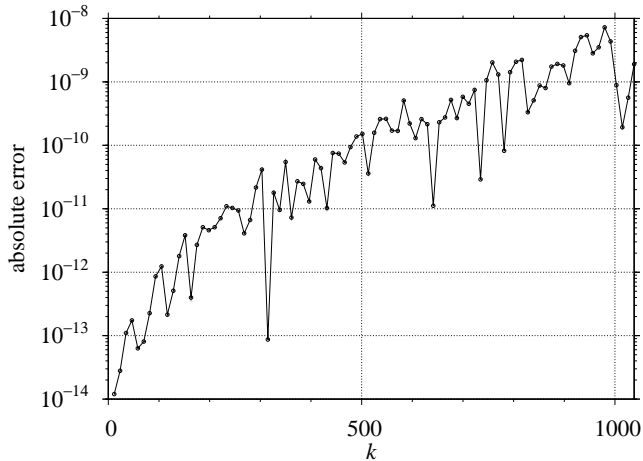}
 }
  \else\drawing 65 10 {soc error of alpha 15 and 35 digit}
 \fi 
 \caption{\label{f:alpha2precisions}
 Discrepancy between  15- and 35-digit calculation of the local
 prefactor  exponent $\alpha_{\rm loc}(k)$ along  $k_2 / k_1 = 5 / 3$,
 as an estimate of the absolute error on $\alpha_{\rm loc}$.
} 
\end{figure}
To estimate the asymptotic value $\alpha_\infty$ we must
\textit{extrapolate} the data beyond the largest available 
wavenumber at which they are known with acceptable accuracy.
Since the only cause of error in $\alpha_{\rm loc}(k)$ are
rounding errors, we can measure such errors by comparing runs having different
levels of precision. Fig.~\ref{f:alpha2precisions} shows the
discrepancy (absolute error) of $\alpha_{\rm loc}(k)$ obtained with 15
and 35 digit precision. The error is seen to grow with the wavenumber
in an approximately exponential fashion, the highest value being 
about $10^{-8}$ around wavenumber 1000. We shall see that the error
involved in the extrapolation may be much larger than $10^{-8}$.

\begin{figure}[ht]
 \iffigs   
  \centerline{%
 \includegraphics[scale=0.7]{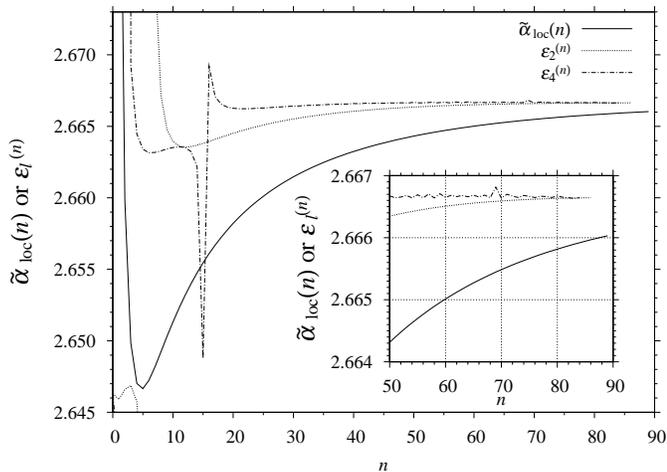}%
 }
 \else\drawing 65 10 {epsilon algorithm}
 \fi  
 \caption{\label{f:epsilon} Local prefactor exponent 
$\tilde \alpha_{\rm loc}(n)$ for the $n$th point along the line $k_2/k_1 =5/3$
which has  $(k_1,\, k_2) =  (5n,\, 3n)$; it is shown together with 
its second and fourth order epsilon-algorithm extrapolated values. 
Inset: enlargement for $n > 50$.  }
\end{figure}

One well-known difficulty with  extrapolation is that the problem may not
be well-posed unless one has 
additional information on the functional form of the convergence  to zero
of the remainder $\alpha_\infty - \alpha_{\rm loc}(k)$. In
Ref.~\cite{shelley}, which deals with the shape of a vortex sheet continued
analytically to complex parameters, it is assumed that  branch singularities
of unknown exponent are present  and that the high-$k$ behavior of the
one-dimensional Fourier transform can be obtained from Laplace's
method to leading and first subleading orders; the inclusion of the
first subleading correction  allows a much
improved determination of the exponent. This extrapolation procedure
is equivalent to assuming that 
the remainder $\alpha_\infty - \alpha_{\rm loc}(k)$ goes 
to zero as $1/k$. For our problem, unfortunately no simple
functional form of the remainder, such as algebraic, exponential or
inverse logarithmic decrease, gives a satisfactory fit. 
An efficient extrapolation method  for a 
wide range of functional behaviors of the remainder is the 
\textit{epsilon algorithm} 
of Wynn \cite{wynn}, related to the Shanks transform method \cite{shanks}.
It is an algorithm for acceleration of convergence of a sequence 
$ S = \left(s^{(0)},\, s^{(1)},\, s^{(2)},\, 
\ldots,\quad  s^{(i)}\right) \in \mathbb C$,
and it comprises the following initialization and iterative phases:\\ 
Initialization: For $n = 0,1,2,\ldots$
\begin{equation}
\varepsilon^{(n)}_{-1} = 0 \quad (\mbox{artificially}), \quad \varepsilon^{(n)}_{0}  = s^{(n)}\;,
\end{equation}
Iteration: For $n = 0,1,2,\ldots$
\begin{equation}
\varepsilon^{(n)}_{l + 1}
= \varepsilon^{(n + 1)}_{l - 1}
  + \left[\varepsilon^{(n + 1)}_{l} - 
\varepsilon^{(n)}_{l}\right]^{-1}\;.
\end{equation}
After a few iterations of the algorithm,  applied to 35-digit SOC data, the
$\varepsilon^{(n)}_{l}$'s with even $l$ become almost
constant and give an estimate of the \textit{extrapolated exponents}
(see Fig.~\ref{f:epsilon}). The epsilon-algorithm
extrapolated exponents will be used when discussing results (unless
otherwise stated). We have also used the recently
introduced asymptotic interpolation method of van~der~Hoeven
\cite{jorasint} which strips off successively leading and subleading
terms by suitable transformations before doing the interpolation. This
method works impressively for the passive scalar model discussed in
Section~\ref{s:linearized} for which both leading and subleading
terms in the high-$k$ expansion can be determined from numerical data.
In the nonlinear case,  the asymptotic interpolation method gives
exponents consistent with those determined by the  epsilon algorithm 
with a relative error of about $10^{-3}$; we have so far 
not been able to determine numerically the functional form of
subleading corrections.  As we shall see in Section~\ref{ss:yplanetheory},
theory tells us that $\alpha$ should not depend on the angle $\theta$.
We suspect that $\theta$-dependent subleading corrections account 
for the slight apparent variation of $\alpha$ with $\theta$, reported
in Section~\ref{ss:SOCresults}.

\subsection{Results for SOC}
\label{ss:SOCresults}

For the SOC, whose initial stream function is $\cos x_1 + \cos 2x_2$,
we now use the method described in Section~\ref{ss:captalpref} to
calculate the prefactor exponent $\alpha(\theta)$, the decrement
$\delta(\theta)$ and the constant $C(\theta)$.  

\begin{figure}[H]
 \iffigs   
  \centerline{%
 \includegraphics[scale=0.7]{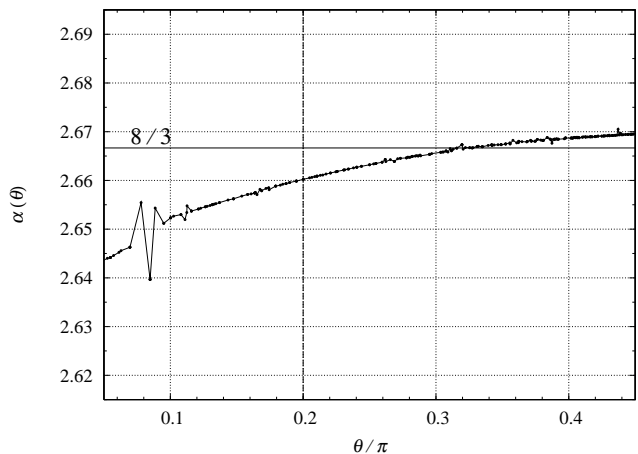}%
 }
  \else\drawing 65 10 {alpha of theta}
 \fi 
 \caption{\label{f:socalpha}Angular dependence of the prefactor exponent
 $\alpha(\theta)$ for SOC (extrapolated by  the epsilon algorithm). 
Below $\theta = 0.2
 \pi$ (long dashed line) the extrapolation cannot be trusted.} 
\end{figure}

\begin{figure}[H]
 \iffigs   
  \centerline{%
 \includegraphics[scale=0.7]{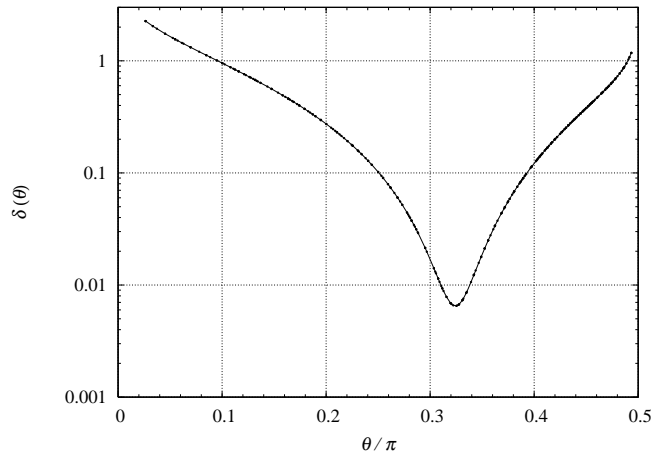}%
 }
  \else\drawing 65 10 {delta of theta}
 \fi 
 \caption{\label{f:socdelta} Angular dependence of the decrement $\delta(\theta)$ 
for SOC.} 
\end{figure}

\begin{figure}[H]
 \iffigs   
 \centerline{%
 \includegraphics[scale=0.7]{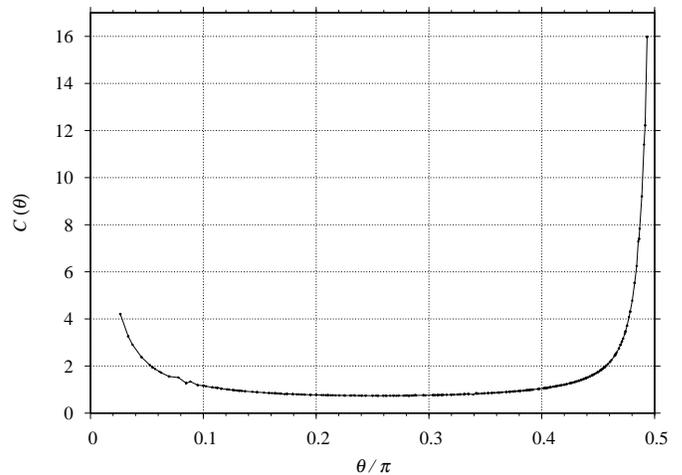}%
 }
  \else\drawing 65 10 {C of theta}
 \fi 
 \caption{\label{f:socC}Angular dependence of the constant $C(\theta)$ 
for SOC.} 
\end{figure}

Figs.~\ref{f:socalpha}, \ref{f:socdelta} and \ref{f:socC} show the
angular variation of $\alpha$, $\delta$ and $C$, respectively,
excluding near-edge ranges where $\theta$ is close to $0$ or $\pi/2$
which deserve separate discussion (see Section~\ref{ss:edgeasymp}).


The most striking result is the very weak angular dependence of the prefactor
exponent, which over the range  $0.2\pi < \theta < 0.45 \pi$ is given by
$\alpha = 2.66 \pm 0.01$, consistent with the theory which predicts
independence on $\theta$ (Section~\ref{ss:yplanetheory}).
This immediately leads to asking if $\alpha_{\rm SOC} = 8/3$. The
short answer is: we do not know. We shall come back to this at length.

The angular dependence of $\delta$ has already been reported in MBF
where it was measured by decomposing the set of directions into small
angular sectors.\footnote{In MBF $\theta$ was varying in the third
quadrant; here, because of the aforementioned change of notation
$\theta$ varies in the first quadrant. Furthermore, the method used in
MBF was less accurate than the present one and there are thus small
discrepancies in the values reported.}  We find that
$\delta(\theta)$ achieves a minimum value $\delta_\star \approx 0.0065$ at
$\theta_\star \approx 0.324\pi$ and it becomes large near the edges. 
In MBF it was reported that the
shell-summed amplitude of $\hat{F} (\k ) $
\begin{equation}
\label{e:ssa}
A (k) \equiv \sum_{k \leq \vert \k \vert \leq k+1 } \vert \hat{F} (\k
) \vert\;,
\end{equation}
a kind of discrete angle average, behaves as $C'k^{-2.16} \ue
^{-\delta_\star k}$ for large $k$. This is consistent with the present
result. Indeed for large $k$, we can evaluate the shell sums
\eqref{e:ssa} by integrating over $kd\theta$ using
\eqref{e:precise_asymptotics} and steepest descent near
$\theta_\star$. This changes the prefactor from $k^{-\alpha}$ to
$k^{-\alpha+1-1/2}\approx k ^{-2.16}$.
 
Finally, $C(\theta )$ is quite flat in the interval $0.1 <\theta <0.4$.

\subsection{Non-universality of the scaling exponent}
\label{ss:nonuniv}

Having established the angle-independence of the prefactor exponent, we now
investigate its dependence on the initial condition. What happens when we
change from SOC  (given by \eqref{proto}) to another initial condition? 
Since 35-digit
computations take up to one month of CPU, we generally used 15-digit accuracy
but there is one important exception (see below). At first we
changed SOC to 
\begin{equation}
\Psi_0(\x) = \cos x_1 + \cos 3x_2\;,
\label{NSOC}
\end{equation}
for which the basic modes in the short-time asymptotics are
$(1,\,0)$ and $(0,\,3)$ between which there is the same  90-degree angle as
for SOC. The prefactor exponent was again indistinguishably close to $8/3$.
For a while this led us to conjecturing the universality of the $8/3$
exponent.
Well \ldots until we tried
\begin{equation}
\Psi_0(\x) = \cos (x_1+x_2) + \cos 2x_2\;,
\label{fortyfive1}
\end{equation}
whose basic modes are $(1,\,1)$ and $(0,\,2)$, forming an angle of 45 degrees.
This gave us an exponent $\alpha \approx 2.54$. 
The same exponent was 
obtained with
\begin{equation}
\Psi_0(\x) =  \cos (x_1+x_2)+ \cos px_2\;,
\label{fortyfive2}
\end{equation}
with $p=1,\,3,\,4$, whose basic modes are different but also form an angle of 45 degrees. We also
did some exploration of the direction dependence of $\alpha$ and, just as for
SOC, did not find any. As we shall see in 
Section~\ref{ss:yplanetheory}, independence on the direction 
can be shown to hold.

\begin{figure}[H]
  \centerline{%
 \includegraphics[scale=0.65]{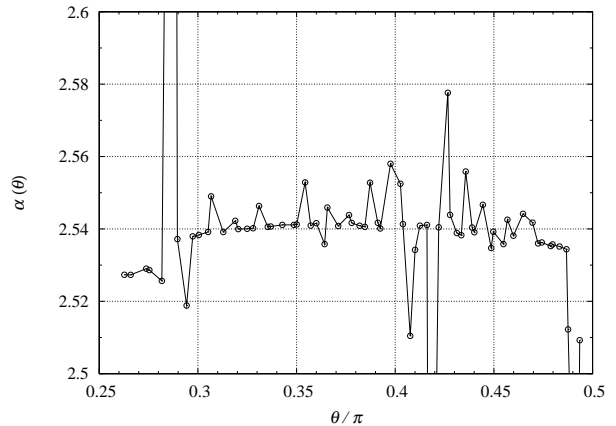}%
 }
 \caption{\label{f:45degalphaoftheta}
Angular dependence of the prefactor exponent $\alpha(\theta)$
for the ``45-degree'' initial condition $\Psi_0(\x) = \cos 2x_1 + 
\cos (x_1+x_2)$
}
\end{figure}

All this was pointing towards non-universality of the prefactor exponent, that
is dependence on the initial condition or at least on the angle between the
basic modes. To ascertain the non-universality we performed a 100-digit 
computation for \eqref{fortyfive1} with $k_{\rm max} =1000$. 
Fig.~\ref{f:45degalphaoftheta}  gives the epsilon-algorithm extrapolated values
of the local prefactor exponent for this calculation as a function of
$\theta$.
Except near the edges the exponent stays very close 
to $2.54$.\footnote{The anomalously low value around 
$\theta/\pi = 0.42$ is caused by a large denominator in the
corresponding slope ($79/21$) which does not permit a reliable
determination of $\alpha$.}
The discrepancy between
$2.54$ and $2.66$ vastly exceeds the estimated
error on the prefactor exponent, as
discussed in Section~\ref{ss:captalpref}. Finally, we report that for
all cases discussed in this section on non-universality, the positivity of
all the Fourier coefficients except one, holds just as for SOC.

\subsection{Intermediate asymptotics near the  edges}
\label{ss:edgeasymp}

In this Section we discuss only SOC, but the theoretical results presented
are easily generalized. 
We have seen that on any line of strictly positive and finite rational slope
the Fourier coefficients decrease exponentially at high $k$ (up to algebraic
prefactors). This is not true for lines of vanishing and infinite slope. 
We can explicitly calculate from the recursion relation 
\eqref{e:difference} all the coefficients having either $k_2=2$ or
$k_1=1$. Indeed along such ``edge lines'' the recursion relations
take the form of first-order linear homogeneous finite difference equations
\begin{eqnarray}
\hat F(k_1,2) &=& \frac{1}{k_1 }\, \frac{k_1^2 -2k_1 + 4}{k_1^2 + 2^2 }\,
\hat{ F} (k_1 - 1, 2)\;, \label{horedge}\\
\hat F(1,k_2) &=& \frac{2}{k_2 } \, \frac{k_2^2 -4k_2 +1}{1+k_2^2 } \, 
\hat{F} (1, k_2 - 2)\;. \label{vertedge}
\end{eqnarray}
At large orders, essentially each coefficient on a horizontal or vertical 
edge line is obtained by dividing by $k_1$ or $k_2/2$ the adjacent lower
order coefficient. Thus they are decreasing roughly as $1/k_1 !$ or
$1/(k_2/2)!$. More precisely, using standard asymptotic methods for
difference equations \cite{BO}, it is easily shown that for
integer $m \to \infty$ 
\begin{equation}
\hat F(m,2) \sim \hat F(1,2m) \sim m^{-5/2} \ue ^{m} m ^{-m}\;,
\label{edgeasympt}
\end{equation}
which decreases faster than exponentially.

If we now consider a ``near edge'' direction with $\theta$ close to $0$ or
to $\pi/2$ we expect that the edge behavior will manifest itself
as intermediate asymptotics making it hard to obtain clean scaling for the
prefactor. We can however easily predict the $\theta$-dependence of
the decrement $\delta$ by the following argument. When $\theta$ is small,
the line through the origin of slope $\tan \theta \approx \theta$ will
intersect the edge $k_2 =2$ at $k_1 \approx 2/\theta$. At this point, by
\eqref{edgeasympt},  the logarithm of  the 
Fourier amplitude is given to leading order by $-(2/\theta) \ln (2/\theta)$.
Assuming that, on the line of slope $\theta$, this point is within the region
of exponential fall-off with decrement $\delta(\theta)$, we obtain
\begin{equation}
-(2/\theta) \ln (2/\theta)\approx -(2/\theta) \delta(\theta)\;,
\label{horapproxdelta}
\end{equation}
which gives (for $\theta \to 0$)
\begin{equation}
\delta(\theta)\approx \ln \left(\frac{2}{\theta}\right)\;.
\label{deltasmalltheta}
\end{equation}
Near the other edge, we obtain by a similar argument  (for $\theta \to \pi/2$)
\begin{equation}
\delta(\theta)\approx \frac{1}{2}\,\ln \left(\frac{1}{\pi -
2\theta}\right)\;.
\label{deltapitheta}
\end{equation}

\begin{figure}%[H]
 \iffigs   
 \centerline{%
\includegraphics[scale=0.55,angle=270]{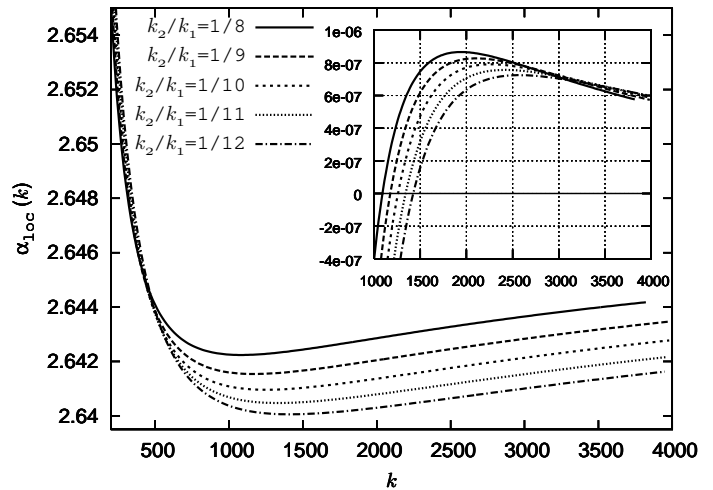}
 }
  \else\drawing 65 10 {edge alpha}
 \fi 
 \caption{\label{f:edgealpha}Wavenumber dependence of local prefactor 
exponent $\alpha_{\rm loc}$ for various small $\theta$. Inset: 
$d\alpha_{\rm loc}(k)/dk$.} 
\end{figure}

We turn now to numerical study of the near edge behavior of Fourier
coefficients. So far we have determined such coefficients in regions having
comparable extensions in the $k_1$ and $k_2$ directions. The structure of the
recursion relation allows us however to determine the coefficients in
rectangular domains having a very small or very large aspect ratio.  We have
seen that the local prefactor exponent behaves non-monotonically with the the
wavenumber when $\theta$ is below a critical value.\footnote{This may be
related to the fact that the global structure seen in Fig.~\ref{f:allinone} is
far from being symmetrical in $y_1$ and $y_2$.}  Consistently, we have found
that for small $\theta$'s a more complex behavior is observed than for
$\theta$'s close to $\pi/2$. We have thus studied the former in more
detail. Because of the slow convergence to asymptotics we need wavenumbers
much larger than in MBF, so we used rectangular domains of size
$4000\times 480$ near $\theta =0$ and of size $200\times 4000$ 
near $\theta =\pi/2$.
Fig.~\ref{f:edgealpha} shows the variation with the
wavenumber of the local prefactor exponent $\alpha_{\rm loc}(k)$ 
for various small $\theta$'s.
It is seen that when $\theta$ decreases, the wavenumber at which
$\alpha_{\rm loc}(k)$ achieves its minimum increases and thus the
extrapolation of $\alpha$ becomes more  difficult.
The situation is much more favorable for the
determination of the decrement, because it is (logarithmically)
large. 
\begin{figure}[H]
\iffigs   
 \centerline{%
 \includegraphics[scale=0.55,angle=270]{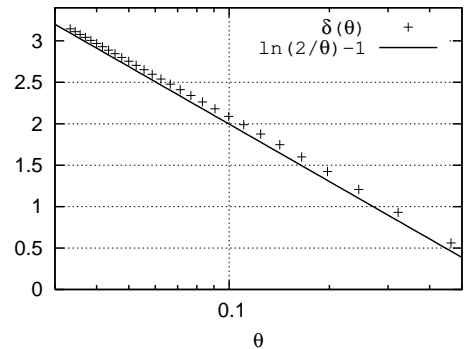}
}
  \else\drawing 65 10 {edge delta}
 \fi 
 \caption{\label{f:edgedelta}Angle dependence of the decrement
   $\delta$
for small $\theta$ in lin-log coordinates (crosses). The continuous line is
the theoretical prediction.} 
\end{figure}
Fig.~\ref{f:edgedelta}  shows the measured decrement together
with a theoretical prediction $\delta(\theta) = \ln (2/\theta)-1$
which includes a subleading correction to the leading-order prediction
\eqref{deltasmalltheta}, obtained by a partially heuristic procedure.
Near $\theta =\pi/2$ the decrement has also logarithmic scaling (not shown),
consistent with the leading-order prediction \eqref{deltapitheta} but
not very clean. As to the constant $C(\theta)$, we found that it
becomes large near the edges. For $\theta\to 0$ the behavior is
roughly $C(\theta) \propto 1/\theta$ but there are substantial
uncertainties
because the constant $C$ is quite sensitive to small errors made on 
$\delta$ and $\alpha$.

\subsection{Beyond short times}
\label{ss:beyondst}

\begin{figure}[t]
 \iffigs   
 \centerline{%
 \includegraphics[scale=1.0]{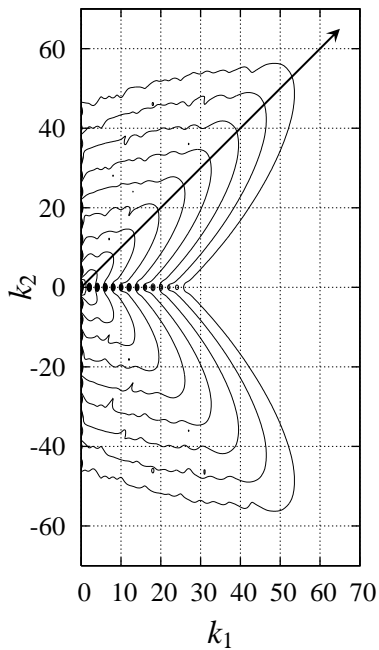}%
 }
  \else\drawing 65 10 {lobe}
 \fi 
 \caption{\label{f:lobe}Contours of the absolute value of the 
Fourier coefficients (logarithmic scale) of the
   stream function at $t = 0.8$ by full spectral simulation for SOC.} 
\end{figure}

\begin{figure}[t]
 \iffigs   
 \centerline{%
 \includegraphics[scale=0.7]{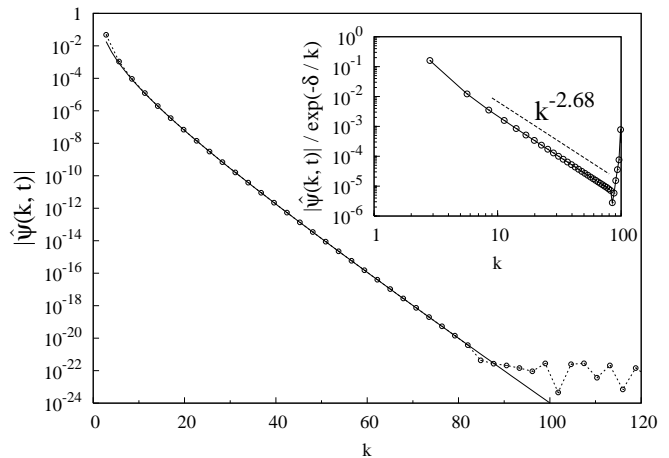}%
 }
  \else\drawing 65 10 { NS t exp}
 \fi 
 \caption{\label{f:nstexp}Absolute value of the Fourier coefficients
 of the stream function at $t=0.8$ along the rational direction 
$k_2/k_1=1$ in lin-log coordinates. A least square fit (continuous line) gives 
$c k^{-2.68} \ue ^{-0.429 k}$. The inset shows the same data 
after division by $\ue ^{-0.429 k}$ in log-log coordinates.} 
\end{figure}

In MBF it was shown, for the SOC initial condition, that deviations
from short-time asymptotics become important around $t=0.1$. More
precisely, deviations from the law $\delta(t) \propto \ln (1/t)$ 
become visible (see
Fig.~2 of MBF). We now investigate numerically the issue of
persistence of the $k^{-2.66}$ law for SOC beyond the time of validity
of short-time asymptotics.  For this we must use a full spectral
simulation with time-marching as in Refs.~\cite{blue,FDR}. A priori
there is no need to use a resolution in excess of
$1024^2$ since we shall see that the $k^{-2.66}$ law deteriorates
significantly after $t=1$. At that time, the decrement 
$\delta \approx 0.4$, which implies that the flow
is extremely well resolved with $1024^2$ modes. Fig.~\ref{f:lobe}
shows the behavior of the absolute value\footnote{Because of the
symmetry of SOC, the Fourier coefficients are real.}  of the Fourier
coefficients of the stream function at $t=0.8$ in the $(k_1,\,
k_2)$-plane (because of the Hermitian symmetry we are not showing negative
$k_1$). It is seen that there is a direction of slowest decrease which
has $k_2/k_1 \approx 1$. At short times the slowest decrease had
$k_2/k_1 \approx 18/11$ but this direction changes in the course of
time. 
Fig.~\ref{f:nstexp} shows the usual exponential decrease with an algebraic 
 prefactor for the  Fourier
amplitude in the direction of slope unity at $t=0.8$. Beyond
  wavenumber 85, rounding errors take over (the calculation has 
15-digit precision).
The same procedure, applied at much later times, e.g. at  $t=1.9$, still
shows some kind of exponential tail but the data are far too wiggly to 
permit the extraction of a reliable power-law prefactor.
We have also repeated the analysis beyond short times
for the flow with initial condition $\cos (x_1+x_2)+\cos 2x_2$, where the basic
modes make an angle of 45 degrees. At time $t=1.4$ the prefactor exponent
is around $2.58$, quite close to the the value $2.54$ reported at short times.

Let us now discuss some of the limitations involved
in the search for prefactor scaling beyond the short-time asymptotics.
We begin with practical limitations. With $1024^2$ modes the direction 
of rational slope $k_2/k_1 =1$ has about 30 points before encountering
15-digit rounding level. Other directions have typically only
ten points and this makes precise determination of the decrement 
$\delta$ and the prefactor exponent $\alpha$ impossible. We thus
cannot comment on any possible angular dependence of
$\alpha$. Calculations with higher resolution require higher
precision in order to lower the rounding noise level and this in turn requires 
enormous computer resources by a  time-marching  full spectral method
if we demand that temporal truncation error be at rounding level. 

There is a more fundamental issue regarding the validity of the
short-time asymptotic r\'egime. For SOC, this r\'egime breaks down 
around $t=0.1$, as far as the temporal behavior
of $\delta(t)$ is concerned. Actually the short-time approximation is
strongly non-uniform with respect to  the wavenumber: high wavenumbers show
discrepancies at much earlier times than $0.1$. For example we know that in the
short-time  r\'egime, all the Fourier coefficients except one are
non-negative, but as early  as $t=10^{-3}$, a 90-digit
calculation by time-marching shows that Fourier coefficients start oscillating
in sign beyond wavenumber forty.\footnote{It matters how precisely we let $t
\to 0$ and $k \to \infty$. For a fixed value of $t$, however small, the
high-$k$ r\'egime discussed in most of this paper may be just an intermediate
asymptotic r\'egime. It is conceivable that the non-universality found here is
confined to this particular asymptotic r\'egime.}. By $t=0.8$ such
oscillations are found in the 15-digit calculation whenever $k_2/k_1 >2$,
irrespective of wavenumber. In the presence of such oscillations, the
functional form we have used in the short-time asymptotics $\propto
k^{-\alpha} \ue ^{-\delta k}$ is clearly invalid.  What is happening has a
geometric interpretation which is more readily understood after reading the
first page of Section~\ref{s:geometry}. In the short-time r\'egime the positivity of
the Fourier coefficients implies that the singular manifold is in the
$y$-plane. Note, however, that in this r\'egime we are ignoring interactions
with Fourier harmonics from quadrants other than the first one since they
only contribute subdominant terms in the short-time expansion of the
hydrodynamic fields. When such terms are taken into account it is likely that
singularities obtained at leading order will be mostly advected by a modified
velocity field which carries the singularities slightly out of the $y$-plane
without changing their nature, as happens in the work of Tanveer and Speziale
\cite{TS93}.  Of course, positivity of the Fourier coefficients will be lost
but not necessarily their scaling properties.  Observe also that the $y$-plane
being a plane of symmetry, this picture implies that there are several pieces
of the singular manifold very close to the $y$-plane. In Fourier space they
produce a kind of interference pattern which at first has very long wavelength
(in $k$). This wavelength becomes shorter and shorter as time advances and the
singular manifold moves further away from the $y$-plane.\footnote{Somewhat
similar interference patterns are obtained when the short-time asymptotics is
extended to the Navier--Stokes equation (with viscosity scaling as $1/t$).}

\section{The geometry of the pseudo-hydrodynamic flow}
\label{s:geometry}

In two-dimensional simulations of hydrodynamics, considerable insight
is usually obtained by looking at flow features in the physical space.
This is much simpler in two dimensions than in three, provided that
the relevant features are in the \textit{real} $\rset^2$ space.  Here
the most important features are in the \textit{complex} $\cset^2$
space, which is equivalent to having four real
dimensions. Fortunately, as explained in Section \ref{s:pshydr}, we
can make use of only two real dimensions by working in the $y$-plane
above $(z_1,\, z_2) = (\pi,\, 0)$ which extends in the (pure)
imaginary directions. As already briefly mentioned in Section~4 of
MBF, the positivity of all the Fourier coefficients $\hat{F}(k_1,
k_2)$ (except $\hat F(1, 0)$) and the exponential decrease with the
wavenumber, imply that the solution to the (short-time asymptotic)
Euler equation has a line of singularities $\mathcal S$ in the $(y_1,
y_2)$-plane. Indeed, since only harmonics with non-negative $k_1$ and
$k_2$ are present, we may rewrite
\eqref{defpsi}-\eqref{psifourierlaplace} as a Taylor series in two
variables
 \begin{eqnarray}
&&\!\!\!\!\!\!\!\!\psi(\y) -\frac{1}{2} y_1+y_2 =
\sum_{k_1=0}^{\infty} \sum_{k_2=0}^{\infty} \hat F(k_1,
k_2)\,\zeta_1^{k_1} \,\zeta_2^{k_2},
\label{doubletaylor}\\
&&\!\!\!\!\!\!\!\!\zeta_1 \equiv \ue^{y_1},\qquad \zeta_2 \equiv
\ue^{y_2}\;.
\label{defzeta1zeta2}
\end{eqnarray}
If we now hold $y_2$ (and thus $\zeta_2$) fixed and sum over $k_2$, we obtain
a Taylor series in $\zeta_1$ such that all its coefficients (except possibly
the first one) are positive. By Vivanti's theorem \cite{Dienes}, if such a
series has a finite radius of convergence (as is the case here because of the
aforementioned exponential decrease), the singularity in the complex
$\zeta_1$-plane nearest to the origin is on the positive real axis at a
location $y_1 =y_1^\star (y_2)$, which depends on $y_2$. The function
$y_1^\star (y_2)$ defines an object which we here call the \textit{singular
manifold} and is the edge of the analyticity domain $y_1 <y_1^\star
(y_2)$.\footnote{More correctly, the singular manifold is a (perhaps analytic)
manifold in ${\mathbb C}^2$ whose intersection with the $y$-plane is
designated here by the same name.} A standard theorem about multi-dimensional
Taylor series states that their domain of convergence is logarithmically
convex (see, e.g., Ref.~\cite{Shabat}).  In our case this just means that the
analyticity domain in the $y$-plane is convex.  As shown in MBF using slightly
different notation, the singular manifold can be constructed either as the
envelope of the family of straight lines $y_1\cos \theta + y_2 \sin \theta =
\delta(\theta)$ (where the decrement $\delta$ has been defined in
Section~\ref{s:fourier}) or as the envelope of analyticity disks.

To numerically construct the pseudo-hydrodynamic
solution in the $y$-plane from the Fourier data we use
\eqref{psifourierlaplace} for the stream function, \eqref{vfourierlaplace} 
for the velocity and \eqref{omegafourierlaplace} for the vorticity.
Although our  Fourier data  typically have
35 decimal digits, it suffices to truncate them to 16 digits to
obtain the various relevant fields in $y$-space with a good accuracy.

\subsection{Presentation of the $y$-plane results}
\label{ss:yplaneresults}

We begin with global
topological features and then turn to a a more local and more quantitative
description.
\begin{figure}[t]
 \iffigs   
  \centerline{%
 \includegraphics[scale=0.9]{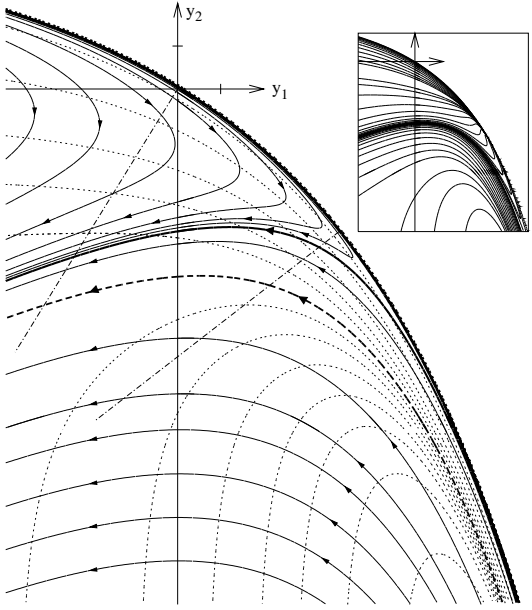}%
 }
  \else\drawing 65 10 { all in one}
 \fi 
 \caption{\label{f:allinone}Global geometry of the flow in the $y$-plane.
 Streamlines (solid lines) and  iso-vorticity lines (thin-dotted lines)
are shown. Thick-solid-crenated line: singular manifold; 
thick-solid line: U-turn separatrix ($\psi \approx 0.5$); 
 thick-dashed line: vorticity separatrix ($\omega =0$ and $\psi = \ln 2$).
The ticks on the two axes correspond to coordinate $0.25$. 
 Inset: Contours of absolute value of the cotangent of the angle between the 
streamlines and 
 the iso-vorticity lines as a measure of depletion of
 nonlinearity.} 
\end{figure}
Fig.~\ref{f:allinone} gives a global view of the flow in the $y$-plane.\footnote{When
magnifying this figure, ADOBE READER$^{\hbox{\scriptsize\textregistered}}$~7
or higher is recommended.} The
outer edge of the flow region, which passes very close to the origin
is the singular manifold. At large distances on the upper
left and the lower right, respectively, the singular manifold has logarithmic
branches. Close to the singular
manifold, the streamlines follow it until they make a U-turn and eventually
plunge into the third quadrant ($y_1<0, \, y_2<0$) where they become straight
with slope $1/2$ at large distances. An important feature is the
U-\textit{turn separatrix}, above which stream lines make U-turns which become
increasingly sharp when moving to the lower right, and below which there are no
U-turns. Vorticity contours starting close to the singular manifold far on the
upper left get pressed increasingly close into the singular manifold
when moving
to the lower right. The \textit{vorticity separatrix} 
divides negative vorticity (above) and positive vorticity
(below).  It approaches the singular manifold in the
lower right but not as fast as the U-turn separatrix. In view of the
Jacobian formulation of the Euler equation, the vorticity
separatrix is clearly also a streamline.  Hence, the strong depletion
of nonlinearity evidenced by accumulation of contour lines near this
separatrix  on the inset of Fig.~\ref{f:allinone}. The depletion is
here
measured by plotting the absolute value of the
cotangent
of the angle between $\nabla \psi$ and $\nabla \omega$.
\begin{figure}[t]
 \iffigs   
  \centerline{%
 \includegraphics[scale=0.8]{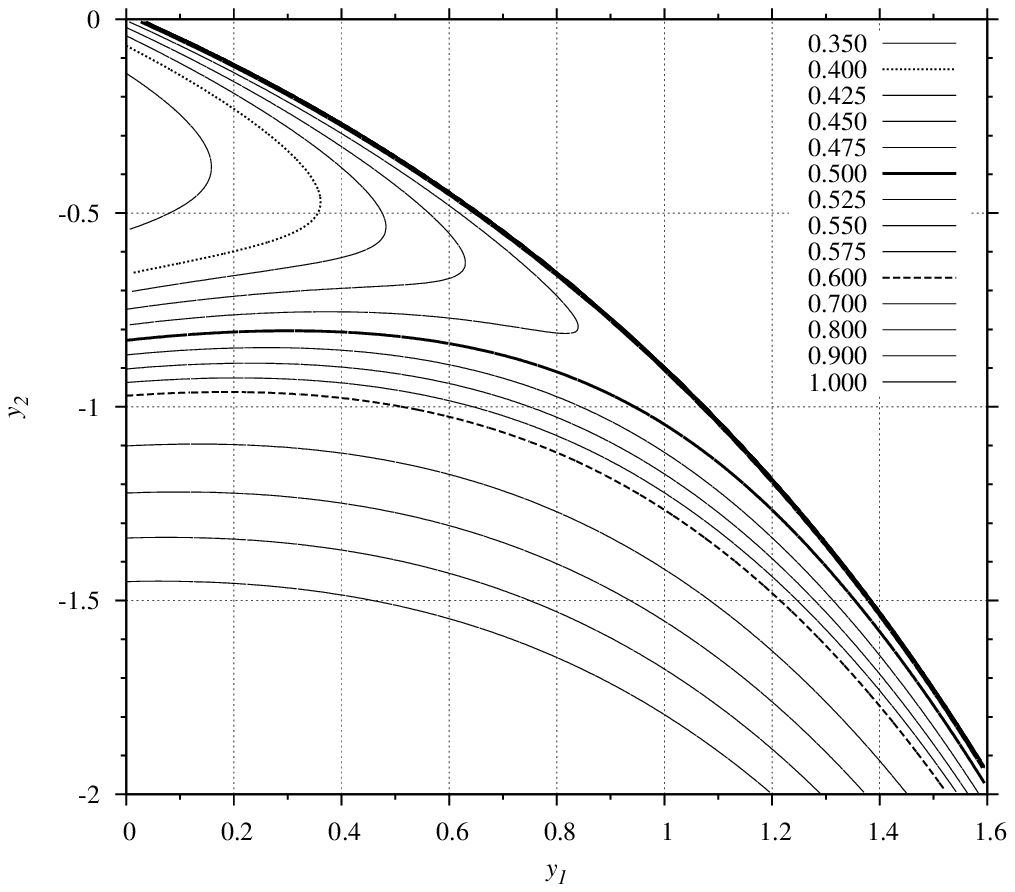}%
 }
 \vspace{3mm}
  \centerline{%
 \includegraphics[scale=0.8]{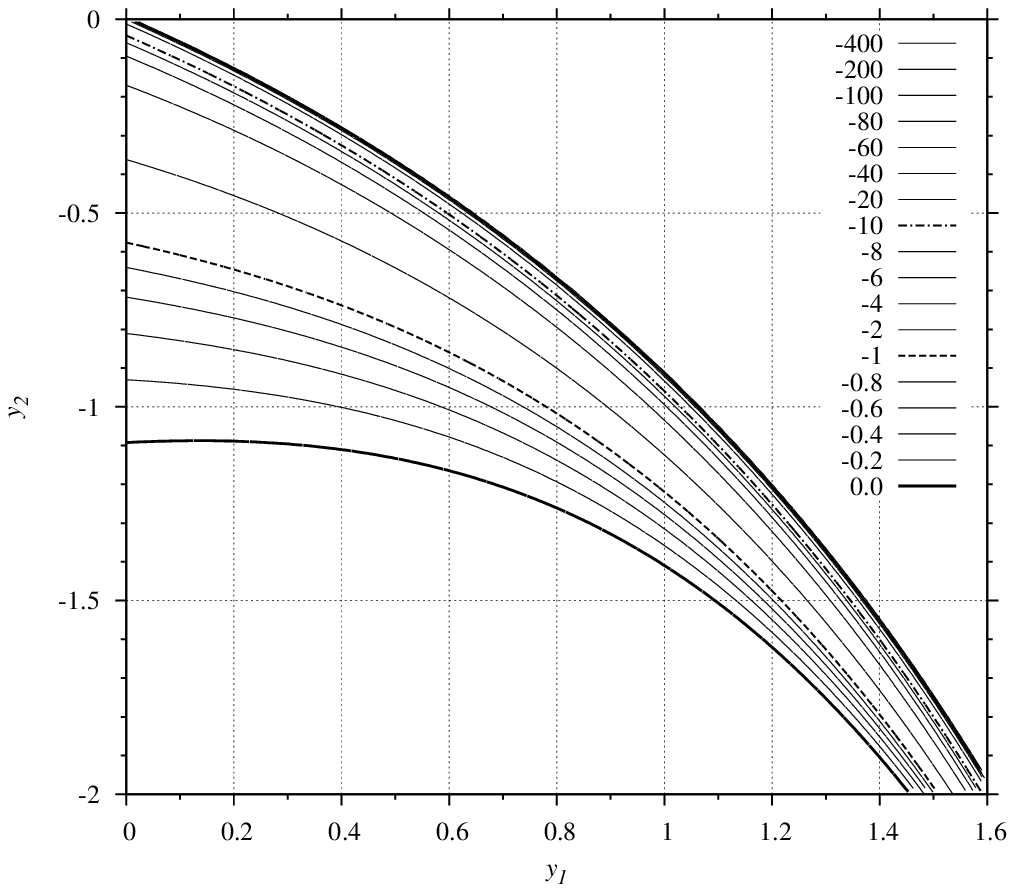}%
 }
  \else\drawing 65 10 { zoom }
 \fi 
 \caption{\label{f:zoom} Enlargements of Fig.~\ref{f:allinone}  around the 
point $(y_1,\, y_2) =    (0.8,\, -1.0)$  showing
the streamlines (upper figure) and the vorticity contours (lower figure).
Only negative vorticity contours are shown.
 } 
\end{figure}

Figs.~\ref{f:zoom}  shows the stream function and the
vorticity with more details in a region of particular interest. Increasingly
sharp U-turns of the stream lines are seen when moving to the lower right into
the narrowing channel separating the singular manifold from the U-turn
separatrix. It is seen that the vorticity becomes very large and negative near
the singular manifold, while the stream function  remains finite with
a value around $\psi = 0.5$, the same as on the U-turn
separatrix. Thus, the singular manifold which is simultaneously a limiting
case of a 
streamline and of a vorticity contour, displays strong depletion of 
nonlinearity as seen on the inset of Fig.~\ref{f:allinone}. We also looked
at the velocity field (not shown);  close to the
singular manifold the velocity is parallel to this manifold and  
decreases in modulus when moving down and to the right.
Note that, contrary to the 
vorticity, the velocity does not grow explosively when approaching
the singular manifold; there is no numerical evidence against the plausible
assumption that the velocity has 
a finite limit on the singular manifold which is tangent to this
manifold.  Similarly, the pressure (not shown) also
appears to have  a finite limit on the singular manifold.
\begin{figure}%[H]
 \iffigs   
  \centerline{%
 \includegraphics[scale=0.65]{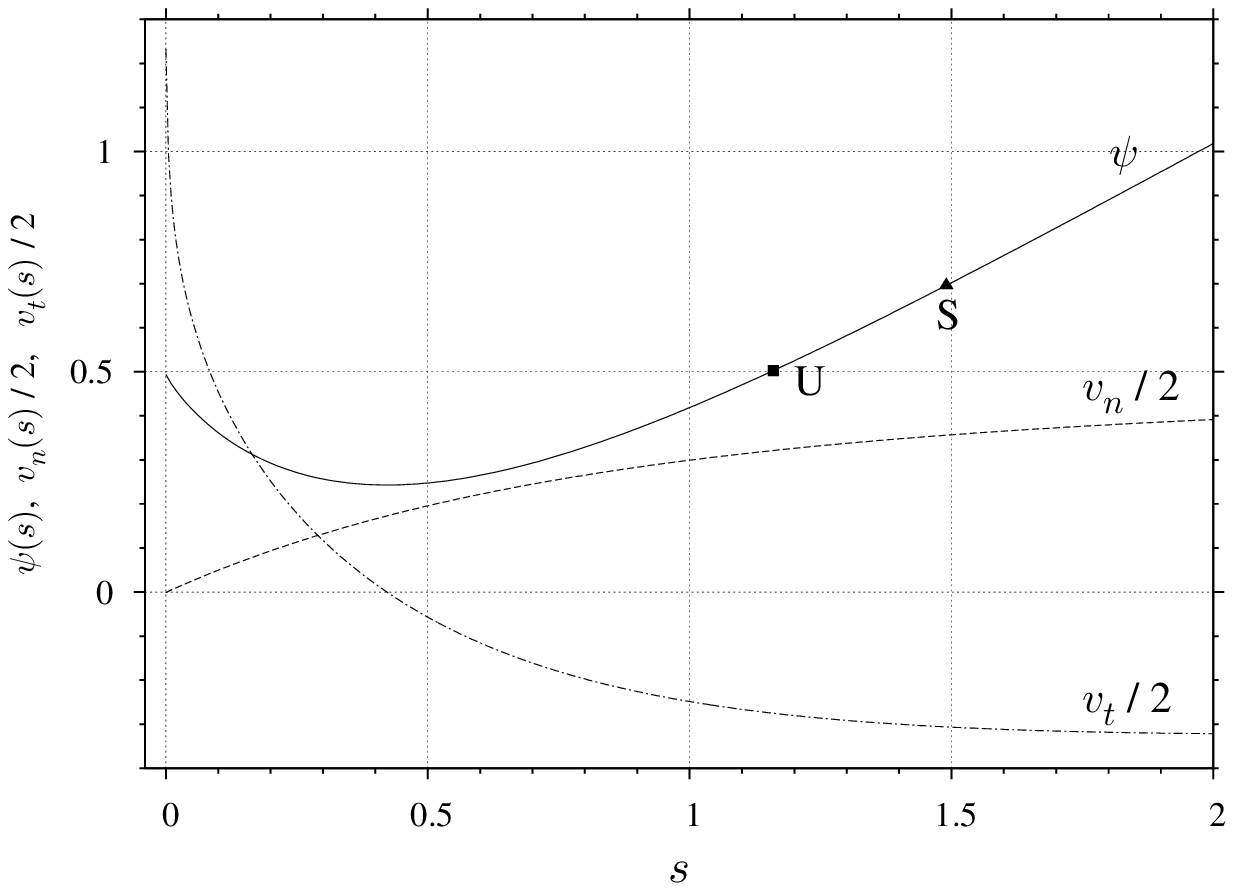}%
 }
  \centerline{%
 \includegraphics[scale=0.7]{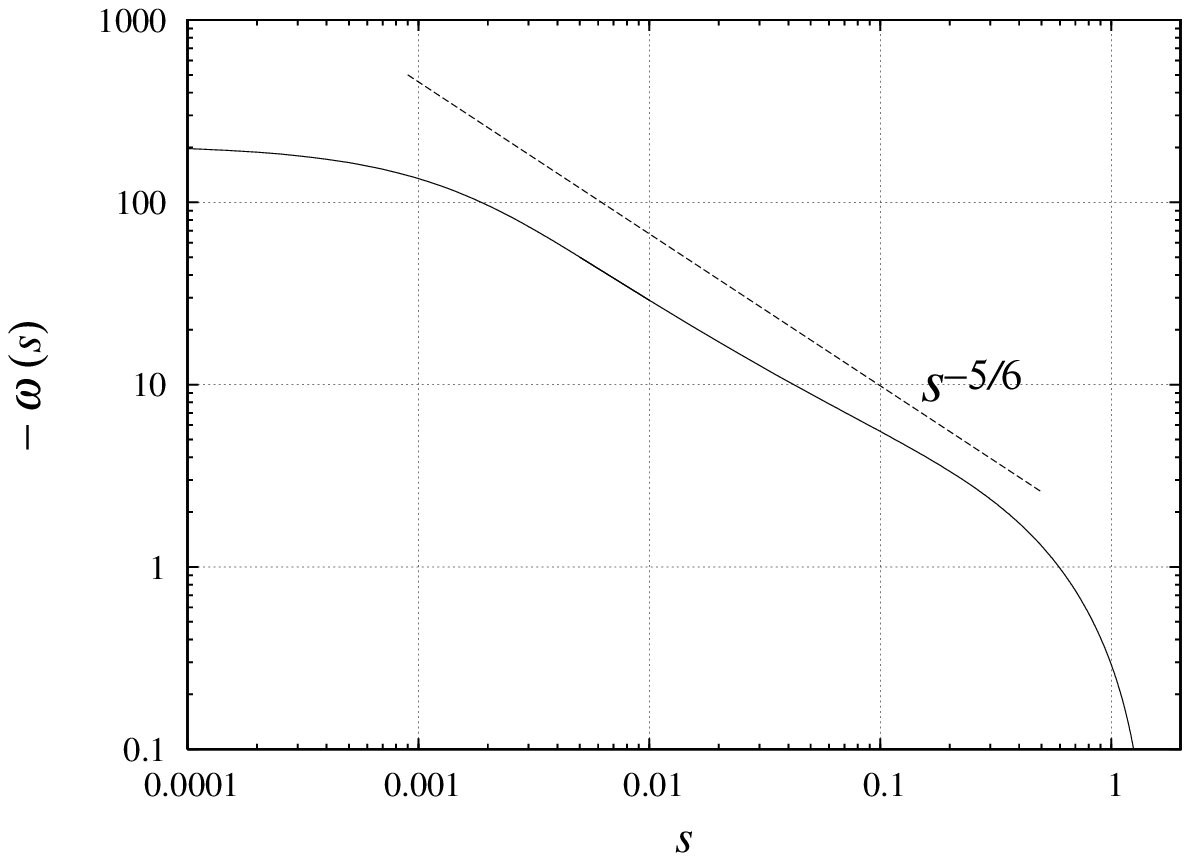}%
 }
  \else\drawing 65 10 { cut 1}
 \fi 
 \caption{\label{f:cut1}Upper figure: stream function $\psi$, velocity 
components (suitably rescaled) $v_n$ and
 $v_t$ normal and parallel to the singular manifold along the line of cut 
normal to the singular manifold passing through the origin, shown as
 a dashed-dotted line on Fig.~\ref{f:allinone}.  
U and  S identify the places where the cut intersects 
the U-turn separatrix and the vorticity separatrix.
Lower figure: negative vorticity along the same cut in log-log
coordinates.}
\end{figure}

For a better quantitative grasp we show in Fig.~\ref{f:cut1} 
a one-dimensional cut of the
two-dimensional fields along the  normal to the singular manifold passing 
through  the origin, shown as a dashed-dotted line on
Fig.~\ref{f:allinone}.  
It is seen that the stream
function takes the finite value $\psi_{\rm sing} \approx 0.5$ on the
singular manifold and the same value at the U-turn separatrix and that the
vorticity follows approximately a power law $-\omega \propto s^{-\beta}$ with
$0.7 <\beta < 0.9$, where $s$ is the distance to the singular manifold.
The scaling is however rather poor; it gets even worse when repeating the
same analysis along the  other  dashed line normal to the singular
manifold, shown on Fig.~\ref{f:allinone}.
It is also seen that at the singular manifold the normal velocity
vanishes linearly.  We now turn to comments and theoretical
explanation of most of these features

\subsection{Bridging $k$-space and $y$-space results}
\label{ss:bridging}

\begin{figure}%[H]
  \centerline{%
 \includegraphics[scale=0.5]{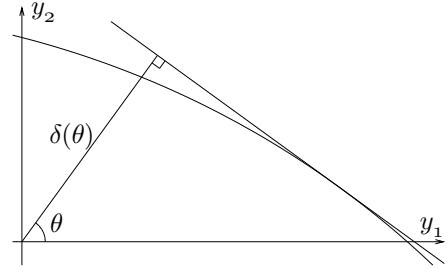}%
 }
 \caption{\label{f:envelope}Construction of the singular manifold
 from the logarithmic decrement $\delta(\theta)$.}
\end{figure}

As we shall now see, it is quite obvious to relate the \textit{leading-order}
asymptotics 
\eqref{e:precise_asymptotics}  of the Fourier coefficients at large $k$
and the \textit{leading-order} behavior near the singular manifold in 
$y$-space. To explain the  poor scaling observed for the vorticity in 
$y$-space, we need to take into account subleading corrections as we shall
also discuss. The
``Fourier--Laplace'' representations \eqref{psifourierlaplace},
\eqref{vfourierlaplace} and \eqref{omegafourierlaplace} for the
(pseudo-hydrodynamic) stream function, the velocity and the vorticity,
connect $k$- and $y$-space functions. 
Consider, for example, the vorticity; using 
\eqref{e:precise_asymptotics} and polar
coordinates  $\k = k(\cos \theta,\, \sin \theta)$ we can rewrite it  as 
\begin{eqnarray}
\omega (\y) &=&  -\sum_{\k}
C(\theta ) k ^{-\alpha+2} \ue ^{ - k h(\theta;\y)}\;,\label{omegapolar}\\ 
h(\theta;\y)&\equiv &\delta (\theta ) - 
y_1 \cos \theta - y_2 \sin \theta\;.
\label{defh}
\end{eqnarray}
The convergence properties at high
wavenumbers of this sum will depend crucially on the sign of the
decrement $h(\theta;\y)$. If
\begin{equation}
\min_\theta \,h(\theta;\y) >0\;,
\label{signexpo}
\end{equation}
all the exponentials are decaying and the sum will be finite. If the
minimum is negative, the sum is divergent. In the borderline case of
a vanishing minimum, the algebraic prefactors will determine
convergence.  If $\delta(\theta)$ is a smooth function of $\theta$, as
our numerical results suggest, the minimum corresponds to a vanishing
derivative with respect to $\theta$. Hence, the borderline case is
characterized by the following two equations:
\begin{eqnarray}
\delta(\theta) -y_1\cos\theta -y_2\sin\theta &=& 0\;, \label{border1}\\
\delta'(\theta)+y_1 \sin\theta -y_2 \cos\theta &=&0\;,
\label{border2}
\end{eqnarray}
where $\delta'(\theta)$ is the derivative of $\delta(\theta)$. Those
points $\y_\star(\theta) =(y_{\star 1},\,y_{\star 2}) $ which satisfy 
\eqref{border1}-\eqref{border2} are on the singular
manifold. Conversely,
$\delta(\theta)$ is the distance from the origin to the tangent 
at the singular manifold which has the slope $\theta -\pi/2$
(see Fig.~\ref{f:envelope}). It follows that the
singular  manifold is the envelope of such lines. 
In MBF this result was derived  by Poincar\'e's pinching argument.

From \eqref{border1}-\eqref{border2} it is easily shown that the
near-edge behavior of $\delta(\theta)$ given by 
\eqref{deltasmalltheta}-\eqref{deltapitheta} implies logarithmic
branches for the singular manifold:
 $y_2 \simeq (1/2) \ln (-y_1)$  for large negative $y_1$ and $y_1
\simeq \ln (-y_2)$ for large negative~$y_2$.  

We observe  that $s \equiv\min_\theta \,h(\theta;\y)$ is the shortest Euclidean
distance of $\y$ to the singular manifold, with a plus sign when $\y$ 
is below the singular manifold and a minus sign when it is above.  
Let us assume that $\y$ is below or on the singular manifold and
let us denote by $\theta_\star(\y)$ the value of $\theta$ where the 
minimum is achieved. Near this  minimum we can Taylor-expand the decrement
\begin{equation}
h(\theta;\y)= s+\frac{1}{2}h_{\star}''\left(\theta -\theta_\star\right)^2
+O\left(\left(\theta -\theta_\star\right)^3\right)\;,
\label{taylorh}
\end{equation}
where $h_{\star}'' \equiv \partial^2 h(\theta_\star,\y)/\partial
\theta^2$. The convergence of the sum \eqref{omegapolar} depends
only on the high-$k$ behavior, where we can, to leading order, replace
the sum by an integral over $kdkd\theta$, to obtain a ``continuous
approximation''
\begin{equation}
\omega _{\rm cont} (\y) = 
 -\int_0^{2\pi } d\theta \int_0^{\infty } dk\,
C(\theta ) k ^{-\alpha+3}\, \ue ^{ - k h(\theta;\y)}\;.
\label{omegacont}
\end{equation}
When $s$ vanishes or is  small and positive, we can evaluate the angular
integral in \eqref{omegacont} by steepest descent
\begin{equation}
\omega_{\rm cont}(\y_\star) \simeq  -C(\theta_\star)\sqrt{\frac{2\pi}
{h_{\star}''}}\int_0^\infty dk k^{-\alpha + 5/2}\,\ue ^{-ks}\;.
\label{asympomegastar}
\end{equation}
On the singular manifold, $s=0$ and it is clear that the integral
over $k$ is ultraviolet-divergent  as soon as $\alpha\ge 3/2$. 
All the values of
the prefactor exponent $\alpha$ considered in this paper are at least
$5/2$ and thus give an \textit{infinite vorticity at the singular manifold}.
The same analysis applied to the stream function and to the velocity
gives ultraviolet-convergent integrals. For small positive $s$, we obtain from
\eqref{asympomegastar}
\begin{eqnarray}
\omega_{\rm cont}(\y_\star) &\simeq&  -C(\theta_\star)\sqrt{\frac{2\pi}
{h_{\star}''}} \,\Gamma(7/2 -\alpha)\, s^{-\beta}\;, 
\label{omegacontas}\\
\beta &=& \frac{7}{2}-\alpha\;,
\label{alphplusbet}
\end{eqnarray}
where $\Gamma(\cdot)$ denotes the Gamma function.

\begin{figure}[H]
  \centerline{%
 \includegraphics[scale=0.70]{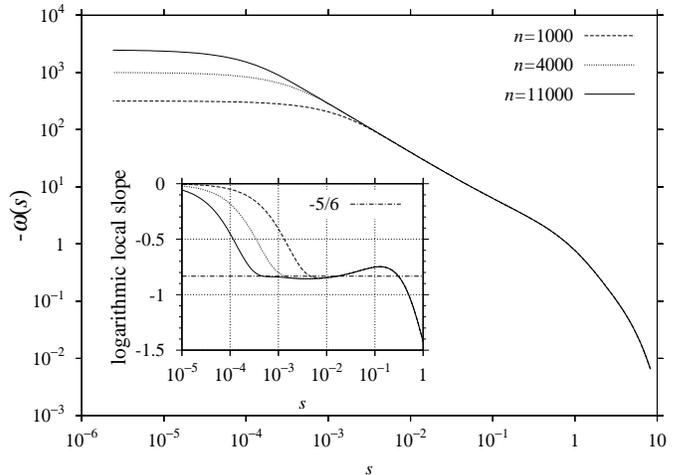}
 }
 \caption{\label{f:synthetic}
Same as lower part of Fig.~\ref{f:cut1} but
for the synthetic data with various values of the  maximum wavenumber $n$.
Inset: corresponding logarithmic local slopes. The predicted scaling exponent
is  $-5/6$.
 }
\end{figure}

For SOC initial conditions, $\alpha \approx 8/3$ and thus the vorticity
diverges to leading order with a  $s^{-5/6}$ law, when approaching the singular
manifold. The subleading corrections causing the poor scaling 
seen in Section~\ref{ss:yplaneresults} are of various sorts. First, there
are subleading corrections to \eqref{e:precise_asymptotics} whose simplest
manifestation is the discrepancy between the local scaling exponent
$\alpha_{\rm loc}(k)$ and
its extrapolated value $\alpha_{\infty}$, as discussed in 
Section~\ref{ss:captalpref}. As already stated, we do not know the 
functional form of
such corrections. Second there are subleading corrections coming from
having approximated the Fourier--Laplace sums by integrals. It is easily
shown that they contribute $O(s^0)$ to the vorticity in $y$-space. Third
there are the subleading corrections to the continuous approximation
\eqref{omegacont}, which may be shown to be $O(s^{-1/3})$. A simple way to
determine how much the scaling is degraded by the second and third
type of corrections is to use\textit{synthetic data} for which the Fourier coefficients are
given exactly by \eqref{e:precise_asymptotics}. It is then easy to change
the resolution 
$n\times n$ and to find out how large $n$ should  be for clean leading-order
scaling to emerge. We performed such a calculation with $C$ and $\alpha$
constant and the values of 
$\delta(\theta)$ taken from the actual Euler SOC data. 
From the synthetic data, using 
\eqref{omegafourierlaplace} we then calculate a synthetic vorticity 
$\omega_{\rm synth}(\y)$ just as in  Section~\ref{ss:yplaneresults}.
Fig.~\ref{f:synthetic} shows  $-\omega_{\rm synth}(\y)$ in log-log
coordinates for three values of $n$. The lowest one $n=1000$ is comparable
to what is used in the actual Euler calculation: the scaling is very
poor. Only when we increase the resolution more than tenfold to $n=11,000$
do we begin to see a clean $s^{-5/6}$ scaling.

\subsection{Theory of $y$-plane pseudo-hydrodynamics}
\label{ss:yplanetheory}

The starting point for theory in the $y$-plane is of course the
pseudo-hydrodynamic vorticity equation and its boundary conditions
far into the third quadrant,  derived in
Section~\ref{s:pshydr} and repeated here for convenience:
\begin{eqnarray}
&&\!\!\!\!\!\!\!\!\!\!\!\!\!\!\v \cdot\nabla \omega +\omega=0,
\label{repseudoomega}\\
&&\!\!\!\!\!\!\!\!\!\!\!\!\!\!\v \simeq \left(-1,\, -{1\over2}\right),
\quad\!\!\! \omega \simeq {1\over2}\ue ^{y_1}
-2\ue ^{2y_2},\quad\!\!\! y_1,\, y_2 \to -\infty.
\label{reboundvomega}
\end{eqnarray}
Alternatively we can rewrite \eqref{repseudoomega} as
$J(\psi,\omega) = \omega$ or as $J(\psi,\ln|\omega|) =
1$. Thus the map from $(y_1,\,y_2)$ to $(\psi,\, \ln|\omega|)$
is area preserving. The vorticity separatrix was defined by
$\omega=0$; so that  that the Jacobian of the stream function and of the
vorticity is zero along this line. Hence it is also a 
streamline.\footnote{By changing $\omega$ into
$1/\omega$ in \eqref{repseudoomega}, we can show similarly that the singular
manifold, at which $1/\omega =0$, is also a streamline.} It
is easily shown that the value of the streamfunction on this line is
$\ln 2$. Indeed, as we follow the vorticity separatrix far into the 
third quadrant, we obtain from \eqref{reboundvomega} that $y_1 \simeq
2y_2 +2\ln 2$. Since $\psi =(1/2) y_1 -y_2$ (up to exponentially small
terms), we obtain the result claimed. We have also checked numerically
that the value of the stream function on the  vorticity separatrix
is $\ln 2$ to at least three decimal places.

Depletion of nonlinearity near the singular manifold prevents
us from using the dynamical equation \eqref{repseudoomega} to
derive the scaling exponent of the singularities by, e.g., balancing
to leading order the two terms in \eqref{repseudoomega}. Nevertheless,
such balancing gives some useful information, such as the vanishing
of the normal component $v_n(s)$ of the velocity near the singular
manifold (for $s\to 0$) and 
the independence on position of the exponent $\beta$
characterizing the divergence of the vorticity. Since $\beta$ and the
prefactor exponent $\alpha$ are related by $\alpha+\beta = 7/2$,
this will establish the independence of $\alpha$ on $\theta$,
which was rather strongly supported by the numerical results 
reported in Section~\ref{s:fourier}. We now derive these results.
In what follows, points $\y_{\star}$ on the singular 
manifold are parameterized by the angle $\theta$ between the
$y_1$-axis and  the outgoing normal.

To show the vanishing of $v_n(s)$ for $s\to 0$, it 
is convenient to use as \textit{local coordinates}
near the singular manifold the angle $\theta$ 
and the distance $s$. We denote by $v_n(s,\theta)$ and 
$v_t(s,\theta)$ the 
components of the velocity along the inward normal and along the
tangent in the direction of increasing $\theta$. For small $s$, to 
leading order, \eqref{repseudoomega} becomes
\begin{equation}
\left[v_n(s,\theta)\partial_s +v_t(s,\theta){1\over R(\theta)}
\partial_\theta \right] \omega(s,\theta) \simeq - \omega(s,\theta)\;,
\label{localomega}
\end{equation}
where $R(\theta)$ is the radius of curvature of the singular
manifold (the arclength is given by $R(\theta)d\theta$).
We now assume that $\omega \propto s^{-\beta}$ (with $\beta >0$).
If $v_n(0,\theta)$ did not vanish, the first term on the l.h.s.
of \eqref{localomega} would be proportional to $s^{-\beta -1}$,
which for small $s$ could not be balanced by any of the other terms
of the equation.  The argument actually implies the stronger result
that the normal velocity $v_n$ cannot vanish more slowly than
$s^1$. The technique of  Section~\ref{ss:bridging} on bridging
$k$-space and $y$-space results can  be used to show that
$\partial v_n/\partial s$
remains finite at the singular manifold, although it has the
same dimension as the vorticity which becomes infinite.\footnote{In
the proof one uses the vanishing of the normal component of
the velocity at the singular manifold, a consequence of the singular
manifold being a streamline.} Thus $v_n$
actually vanishes linearly with $s$. 

\begin{figure}%[H]
  \centerline{%
 \includegraphics[scale=0.43]{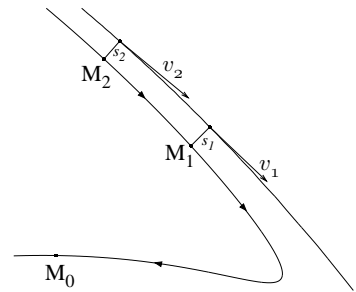}%
 }
 \caption{\label{f:streamsketch}A typical streamline passing near the
singular manifold and performing a U-turn. Although this figure uses
the actual data rather then being a sketch, the distance between the streamline and the singular manifold
has
been somewhat increased for legibility.}
\end{figure}

\begin{figure}[H]
  \centerline{%
 \includegraphics[scale=0.65]{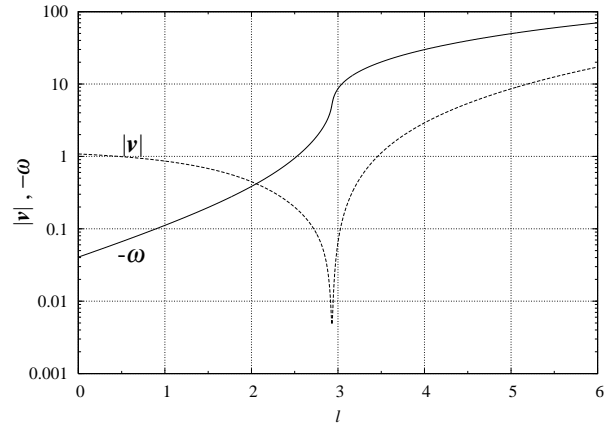}%
 }
 \caption{\label{f:omegavonstream}
 Modulus of the velocity and  vorticity (changed sign) vs arclength $\ell$
 along the  streamline shown in Fig.~\ref{f:streamsketch} which has 
$\psi \approx 0.47$. The point M$_0$ is taken near $\ell = 2.5$;
  the U-turn, M$_1$ and  M$_2$ are near  $\ell = 3$, $\ell = 3.2$, 
$\ell = 3.5$, respectively.
 }
\end{figure}

For the independence of $\beta$ on $\theta$, we integrate 
\eqref{repseudoomega} along a typical streamline passing near
the singular manifold between two points $M_0$ (far from the
singular manifold) and $M_1$ (within a small distance $s_1$), 
so that there is a
U-turn in between which is assumed not to be close to either
$M_0$ or $M_1$ (see Fig.~\ref{f:streamsketch}). 
We obtain
\begin{equation}
\frac{\omega(M_1)}{\omega(M_0)} = \exp \left\{ \int_{M_0}^{M_1} 
\frac{d\ell}{|\v(\ell)|}\right\}\;.
\label{vortint}
\end{equation}
The streamline is here parameterized  by the arclength $\ell$
measured from an arbitrary reference point and growing when moving
into the upper far left,  opposite to the direction of the 
velocity. When moving from $M_0$ to
$M_1$ the smallest velocities and thus the leading-order contribution
to the integral  $\int_{M_0}^{M_1} 
d\ell/|\v(\ell)|$ are expected to come from the immediate neighborhood of
the U-turn. We have checked this conjecture numerically by calculating
the vorticity and the modulus of the velocity along a streamline
chosen to have a rather sharp but well-resolved U turn (see
Fig.~\ref{f:omegavonstream}).
We assume now that the vorticity near the singular
manifold is given to the leading order by $\omega = 
c(\theta) s^{-\beta(\theta)}$, where we temporarily leave the
possibility
that the exponent $\beta$ depends on the parameter $\theta$
associated to the nearest point on the singular manifold. We take
a second point $M_2$ on the same streamline  but further away
from the U-turn. We then have (to leading order)
\begin{eqnarray}
\omega(M_1) &\simeq& c(\theta_1) s_1 ^{-\beta(\theta_1)}\;, \label{M1}\\
\omega(M_2) &\simeq& c(\theta_2) s_2 ^{-\beta(\theta_2)}\;, \label{M2}
\end{eqnarray}
where $(s_1,\,\theta_1)$ and $(s_2,\,\theta_2)$ are the local
coordinates for $M_1$ and $M_2$. From \eqref{vortint}, applied
successively to $M_1$ and $M_2$ we find that
\begin{equation}
\frac{\omega(M_2)}{\omega(M_1)} \simeq  \exp\left\{\int_{M_1}^{M_2} \frac{d\ell}{|\v(\ell)|}\right\}\;.
\label{M1M2}
\end{equation}
Between $M_1$ and $M_2$ the streamline is close to the
singular manifold and the velocity is dominated by its tangential
component; hence we can replace the  r.h.s. of \eqref{M1M2} by
an integral along the singular manifold and obtain to leading order 
\begin{equation}
\frac{\omega(M_2)}{\omega(M_1)}\simeq K_{12} \equiv
\exp\left\{\int_{\theta_1}
^{\theta_2} d\theta\,\frac{R(\theta)}{v_t(\theta)}\right\} \;,
\label{M1M2appr}
\end{equation} 
which depends neither on $s_1$ nor on $s_2$. 
We now observe that the solenoidal
character of the velocity implies (again to leading order)
\begin{equation}
s_1v_t(\theta_1)\simeq s_2 v_t(\theta_2)\;.
\label{s1v1s2v2}
\end{equation}
It follows from \eqref{M1M2appr} and \eqref{s1v1s2v2} 
\begin{equation}
\omega(M_2)\simeq c_2\left[\frac{s_1 v_1}{v_2}\right]^{-\beta(\theta_2)}
\simeq K_{12}\,c_1 s_1^{-\beta(\theta_1)}\;.
\label{almostthere}
\end{equation}
Comparison of the middle and the rightmost members gives
\begin{equation}
\beta(\theta_1) =\beta(\theta_2), \quad c_2 =K_{12}\, 
c_1\left[\frac{v_1}{v_2}\right]^{\beta}\;.
\label{done}
\end{equation}
This establishes the independence of the vorticity scaling exponent 
on $\theta$.

\section{A passive scalar model}
\label{s:linearized}

As shown in Ref.~\cite{PM} simple advection of a passive scalar 
by a prescribed velocity field
with just a few Fourier harmonics can easily lead to singularities
because fluid particles may come from or go to (complex) infinity in
a finite time. In the present context of short-time asymptotics, the
equivalent of a passive scalar model is to treat the
(pseudo-hydrodynamic) vorticity 
$\omega$ in \eqref{pseudoomega} as a passive scalar advected by a
prescribed velocity. The simplest prescribed velocity we can take is 
\begin{equation}
\v_{\rm P}(\y) = \left(-1,\, -\frac{1}{2}\right ) +\left(\ue ^{2y_2},\,
\frac{1}{2}\ue ^{y_1}\right)\;,
\label{defvlinearized}
\end{equation}
obtained from the stream function 
\begin{equation}
\psi_{\rm P} \equiv \frac{1}{2}y_1 -y_2 -\frac{1}{2} \ue
^{y_1}+\frac{1}{2} \ue ^{2y_2}\;.
\label{defstreamlinearized}
\end{equation}
This velocity field includes the drift $(-1,\, -1/2)$ resulting from
the shifts of the original coordinates by terms proportional to $\ln
t$ and the contributions from the basic modes.
For our passive scalar model we use the vorticity equation
\eqref{pseudoomega} with the inclusion of an inhomogeneous term whose
precise form does not matter (as long as it does not have itself any
singularity): the singularities  of the passive vorticity stem solely
from advection. Specifically, the passive scalar model
is defined by
\begin{equation}
\v_{\rm P}\cdot \nabla \omega +\omega = 
\frac{3}{2}\ue ^{y_1+2y_2}\;,
\label{linearizedmodel}
\end{equation}
where the r.h.s. is taken to be the interaction term of
the two basic modes.
It is easy to write down Fourier-space recursion relations for this
model and to show that all the Fourier coefficients 
are positive. Numerical solution of the recursion relations gives the
usual type of scaling with a very clean prefactor exponent 
$\alpha = 5/2$.\footnote{With the already cited asymptotic
interpolation method \cite{jorasint} the exponent $\alpha$ is found to differ 
from $5/2$ by less than $10^{-11}$.}
In the $y$-space this implies a blow up of the vorticity $\omega
\propto s^{-1}$ as function of the distance $s$ to the singular manifold.

\begin{figure}%[H]
  \centerline{%
\includegraphics[scale=0.55,angle=270]{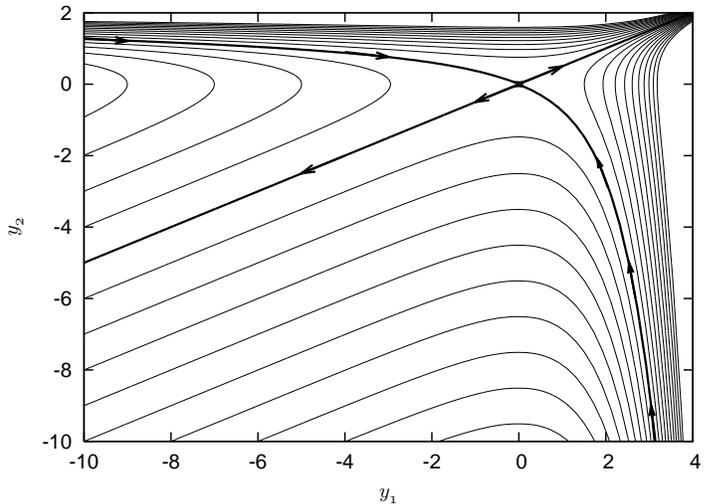}
 }
 \caption{\label{f:streamlinearized}
Streamlines in the $y$ plane for the passive scalar model given by 
\eqref{defstreamlinearized}, which has a hyperbolic stagnation point
at the origin. Thick line with arrows
pointing to the origin: stable manifold of the associated dynamical system
\eqref{dynsystlinearized} which is also the singular manifold for the
vorticity. Thick line with arrows pointing away from  the origin: unstable
manifold. 
 }
\end{figure}

Actually all these results can be derived in a rather straightforward
manner by working in the $y$-space, as we now explain. 
Eq.~\eqref{linearizedmodel} can be integrated along the
characteristics. For this we consider the conservative dynamical
system of fluid particle trajectories in the velocity field $\v_{\rm P}$:
\begin{equation}
\frac{d}{d\tau}\y = \v_{\rm P}(\y)\;.
\label{dynsystlinearized}
\end{equation}
The integral lines are the lines $\psi_{\rm P} = {\rm const.}$, which
are shown in Fig.~\ref{f:streamlinearized}.
At the origin there is a hyperbolic stagnation point near which we have 
$\psi_{\rm P} = -(1/4)y_1^2 +y_2^2 +O(|\y|^3)$. The associated unstable
manifold is simply $y_1=2y_2$, while the stable manifold is the other
solution to $\psi_{\rm P}=0$. 

This hyperbolic stagnation point at the origin completely determines
the scaling of the vorticity. Indeed, in 
Section~\ref{ss:yplanetheory} we derived from the vorticity
equation \eqref{pseudoomega} an expression \eqref{vortint} which shows that
large vorticities stem from low-velocity regions. This derivation, which
did not make use of the fact that the vorticity is the curl of the velocity,
remains valid for the passive scalar model (except for minor changes due
to the presence of an inhomogeneous term). In the fully nonlinear case, the
low velocities were due to the increasingly sharp U-turns 
described in Section~\ref{ss:yplaneresults}. In the present much simpler
case, they are just due to the passage near the hyperbolic stagnation point.
As we follow a streamline upstream (i.e.\ to increasing arclengths $\ell$ 
in the 
notation of Section~\ref{ss:yplanetheory}) we  come closer and closer
to the stable manifold. The latter thus plays the role of the singular
manifold. By the same argument as used in Section~\ref{ss:yplanetheory} the
scaling of the vorticity near the stable/singular manifold is the same
everywhere. It suffices to determine it locally near the stagnation point. 
One way is to parameterize the streamline by the distance $s$ to the stable
manifold (more precisely to its tangent at the origin, since we are doing
a local analysis). By \eqref{vortint}, the growth of the vorticity is 
controlled by 
\begin{equation}
\exp\left\{\int^{\ell} \frac{d\ell'}{|\v(\ell')|}\right\} = \exp\left\{\int^{s}\frac{ds'}{v_n(s')}\right\}\;,
\label{fromelltos}
\end{equation}
where $v_n(s)$ is the velocity component along the (inward) normal $\n =
(-1/\sqrt5,\, -2/\sqrt5)$ to the stable manifold at the origin. 
Near a hyperbolic
stagnation point we have $v_n(s) = \lambda s + O(s^2)$, where $\lambda$ is
the positive eigenvalue of the velocity gradient $\partial_i v_j$ at the
stagnation point. Here, it is elementary to
show that $\lambda =1$. Using this in \eqref{fromelltos} we find
that the vorticity $\omega \propto s^{-1}$, as claimed.
This argument is easily adapted to the  passive scalar models  for
other two-mode initial conditions,
such as those discussed in Section~\ref{ss:nonuniv}. The same $s^{-1}$
behavior is always obtained,
which is thus universal, contrary to what happens in the full nonlinear
case. 
  
Although the  passive scalar model does not predict the exact and non-universal
character of the vorticity blow up for the full nonlinear problem, the
singular manifold is given quite accurately by  the passive scalar model.
In particular it is immediately checked that both have the same logarithmic
branches (at least to leading order). Furthermore in the passive
scalar model the 
stable/singular manifold goes exactly through the origin while in 
the full nonlinear problem it passes within a distance $\delta \approx0.0065$,
as shown in MBF. It may be that such agreements are due to the presence
of very strong depletion of nonlinearity in the full problem, thereby
making a simple linear advection model quite relevant. 

Actually, the passive scalar model can be systematically improved by
enriching the prescribed velocity field through addition of 
higher-order modes. A simple way to do this is to take all
the Fourier modes such that $k_1+k_2/2 \le n+1$. For SOC, we have studied these
``enriched'' passive scalar models for various values of $n$. They all possess
a hyperbolic stagnation point. The associated  positive eigenvalue $\lambda_n$
becomes larger than unity when $n\ge 1$.  The first few values 
for the corresponding prefactor
exponent $\alpha_n = 7/2 -1/\lambda_n$ are:
$\alpha_0 =2.5$, $\alpha_1 \approx 2.594$, $\alpha_2 \approx 2.613$. 
For larger values of $n$
the  growth is very slow; for example $\alpha_{20}\approx  2.618$. 
We also observed that as $n$ grows,
the stagnation point moves to the right and down and the angle between its
stable and unstable manifolds decreases. It is likely that for $n\to \infty$, 
the stagnation point is pushed to
infinity in such a way that its stable and unstable manifold tend
to the singular manifold and to the U-turn separatrix for  the
nonlinear problem, while $\alpha_n \to \alpha$, but the convergence may be 
slow.

\section{Conclusion}
\label{s:conclusion}

The present paper, like Refs.~\cite{blue} and MBF, is mainly concerned
with the short-time asymptotics of the 2D Euler equation in situations
where complex-space singularities are born at infinity at time $t=0+$.
Let us first summarize the main findings of this work, which uses
a mixture of ultra-high precision computations (with up to 100-digit
accuracy) and of theory. Our 
work is specifically concerned with initial conditions
in the form of a trigonometric polynomial;   it is shown in the 
Appendix that this problem can generically be
reduced to one with only two modes. A very
detailed description of the complex singularities is given. For all cases
studied, the Fourier coefficients except one are found to be non-negative (this
was already reported for SOC in MBF). In any  direction of rational slope $\tan
\theta$ not too close to the edges of the Fourier domain, 
the  coefficients of the stream function converge very quickly 
with increasing wavenumbers $k$ to the form  $C(\theta) k ^{-\alpha} \ue ^{-k
  \delta(\theta)}$. The prefactor exponent $\alpha$, determined
with better than one percent accuracy,  is independent of
$\theta$ but is not universal: when the initial modes are orthogonal,
it is indistinguishable from $8/3 \approx 2.66$, whereas with a 45 degree angle
between the initial modes it takes the value $2.54$. We cannot rule 
out that $\alpha$ depends also on the moduli of the initial modes but
we have no evidence that it does.

 It is shown that
the singularity problem can be reformulated as an ordinary steady-state 
(pseudo)hydrodynamic problem in a suitable $y$-plane corresponding to
pure imaginary coordinates. The complex singularities are in this
$y$-plane on a smooth (possibly analytic) curve extending to infinity
with logarithmic branches. The
vorticity diverges as $s^{-\beta}$, where $s$ is the distance to the
singular manifold and $\alpha+\beta =7/2$. We give a full description
of the geometry of streamlines and vorticity contours in the
$y$-plane (Fig.~\ref{f:allinone}). Increasingly sharp U-turns of the 
streamlines near the lower logarithmic branch of the singular manifold
give rise to the vorticity scaling. Very strong depletion of nonlinearity
near the singular manifold prevents application of dominant balance
to determine the scaling exponent of singularities and is likely to be
the reason for the very unusual non-universality of the
singularities. Finally it is  shown that the scaling behavior of
the prefactor persists in time significantly beyond the validity
of the short-time asymptotics, at least as intermediate asymptotics.
However, we do not know if the non-universality of the singularities
found in the short-time r\'egime carries over to the full Euler equation.

The main theoretical shortcomings of this work are our inability so far
to prove the positivity of Fourier coefficients and to derive the
prefactor exponent $\alpha$ (or the vorticity divergence exponent
$\beta$) from the initial conditions (we also failed to identify the nature of
subleading corrections to \eqref{e:precise_asymptotics}). 
We have nevertheless gained some
qualitative understanding with the passive scalar model of 
Section~\ref{s:linearized} that ignores the back reaction of the 
vorticity on the velocity but which sheds interesting light on the
mechanism for producing singularities.
In this toy model,  scaling is controlled 
by a stagnation point of the velocity field, whereas in the full
nonlinear problem the stagnation point is rejected to infinity.

We have described our findings 
in some detail, hoping that
colleagues will be able to help us with the missing theory.

In principle the methods used for the 2D short-time Euler problem can be
extended to various other short-time problems.  One instance is the short-time
r\'egime for the 2D ideal incompressible MHD equations. A preliminary study
for this case indicates that the positivity result does not survive: Fourier
amplitudes display oscillations revealing a richer geometry of the singular
manifold which can no more be captured in terms of just the imaginary
coordinates $\y$. In mathematical terms, one has to
study the amoeba and coamoeba of the singular manifold.\footnote{In
$d$-dimensional algebraic geometry one deals with an algebraic manifold in
complex coordinates $\zeta_1,\ldots \zeta_d$ and the amoeba is defined as the
image of the manifold under the map $\zeta_1 \mapsto y_1 \equiv \ln
|\zeta_1|,\ldots, \zeta_d \mapsto y_d \equiv \ln |\zeta_d|$. The coamoeba is
similarly defined in terms of the argument functions of the $\zeta$'s. The
complex exponentials $\ue ^{-\ui z_1},\ldots,\ue ^{-\ui z_d}$, play here the
role of the $\zeta_i$'s. The name amoeba has been proposed by Gelfand,
Kapranov and Zelevinsky \cite{GKZ} because amoebae sometimes have pseudopods
resembling those of microscopic protozoa. Coamoebae have been introduced by
Tsikh and Passare (private communication).} Oscillations
can be handled by techniques similar to those discussed here, as has already
been done in Ref.~\cite{caflisch}.

Another natural extension of our study is to the 3D Euler equations which also
have a short-time r\'egime.  This is rather straightforward.  A
direct extension of the algorithm used in two dimension requires CPU resources
(time complexity) proportional to $k_{\rm max}^6$ instead of $k_{\rm
max}^4$. This becomes prohibitively large when $k_{\rm max}$ exceeds a few
hundred.  In principle, the time complexity can be reduced to $k_{\rm max}^3$
(with logarithmic corrections) by using FFT's and the recent technique of
``relaxed multiplications'' \cite{Hoeven}.  However in the calculations
reported in MBF and the present paper the magnitude of Fourier coefficients
can vary by several hundred orders of magnitude; this requires special
precautions when applying FFT's unless one is prepared to use several hundred
digits.

We remind the reader that our long term goal is to find out about blow
up in three dimensions (3D). We hope this will not take another 250
years. Progress may however be painfully slow if, as we expect,
numerical experimentation is to play an important part. Indeed, the
amazingly fast growth of computer power observed over the last 50
years, becomes much less spectacular when translated in terms of
resolution achievable in 3D simulations.\footnote{At the moment the
highest resolution accessible in 15-digit precision for
three-dimensional flow without any special symmetries is $2048^3$
\cite{IUYSIK,KIYIU}}. As more powerful computers become available for
investigation of 3D blow up, it would not be advisable to use the new
resources exclusively for increasing the spatial
resolution. Experience on the advantage of ultra-high precision for
singularity studies from the work of Krasny \cite{krasny}, of Shelley
\cite{shelley}, of Caflisch \cite{caflisch} and also from our own work
suggest that it is not safe to use less than 30--35 digits.  Using
flows with symmetries such as the Taylor--Green \cite{TG,BMONMF} or
the Kida--Pelz \cite{Kida,Pelz,CB} flow to boost the resolution
introduces a possible element of non-genericity, but we can always use
such flows to sharpen our tools and then, as computers become more
powerful, turn to flows without symmetry.

Have the results reported in MBF and the
present paper brought us closer to this Holy Grail of 3D blow-up? In a
direct way, we cannot infer anything regarding 3D real blow-up 
from a 2D study of complex singularities at short times. We have
however learned that in this rather restricted framework,
singularities are located on very smooth objects (possibly analytic
manifolds); because the fastest spatial variation is then in the
direction perpendicular to the singular manifold, the  singularities
have strongly depleted nonlinearity in a suitable frame. 
We have already good evidence
that in 2D this smoothness property is not limited to the short-time r\'egime
\cite{blue}. In 3D such a property would be both a curse, since 
dominant balance cannot be used, and perhaps a blessing, since it might
well slow down (indefinitely?) the approach  of singularities to the 
real domain.

\vspace*{4mm}
\par\noindent {\bf Acknowledgments}

We are grateful to M.~Blank, H.~Frisch, J.~van~der~Hoeven, D.~Mitra, A.~Pumir,
A.~Sobolevski\u{\i}, A.K.~Tsikh, P.~Zimmermann and an anonymous
referee for useful discussions and comments.
Part of this work was done while the authors participated in ``Frontiers of
Non Linear Physics'' (Nizhny Novgorod, Russia, July 5--12, 2004) and in
``Singularities, coherent structures and their role in intermittent
turbulence'' (Warwick, UK, September 9--17, 2005). TM and JB were supported by
the Grant-in-Aid for the 21st Century COE ``Center for Diversity and
Universality in Physics'' from the Japanese Ministry of Education. TM was
supported by the Japanese Ministry of Education Grant-in-Aid for Young
Scientists [(B), 15740237, 2003] and by the French Ministry of Education. UF
was supported by the Grant-in-Aid for the 21st Century COE ``Center of
Excellence for Research and Education on Complex Functional Mechanical
Systems'' from the Japanese Ministry of Education. WP had partial 
support from the European network EU RTN no.~HPRN-CT-2002-00282  'Hyk\'e. 
JB, UF and WP had partial support from the F\'ed\'eration de Recherche
Wolfgang Doeblin (FR~2800 CNRS). 
Part of the computational
resources were provided by the Yukawa Institute for Theoretical Physics
(Kyoto) and by the M\'esocentre SIGAMM (Nice).

\vspace{3mm}

\appendix
\renewcommand{\theequation}{\thesection.\arabic{equation}}
\section{Reduction of multimode initial conditions}
\setcounter{equation}{0}
\label{a:multimode}

Here we shall show that the short-time asymptotics of two-dimensional 
Euler flows with
generic initial conditions of trigonometric polynomial type can be
reduced to the study of two-mode initial conditions.  In
Section~\ref{s:pshydr} we have seen that with  two
initial modes $\twop $ and $\twoq $ the
behavior of the stream function
$\Psi (\z ,t)$ for large imaginary arguments $\vert y_1 \vert$, $\vert y_2
\vert $ can be described by the {\it similarity ansatz}
\eqref{simansatz}-\eqref{deftz}. It relies on the fact that
when $\y $ is such that $\pf _2 / \pf _1 \leq y_2 / y_1 \leq \qf _2 / \qf
_1 $ the leading-order factors accompanying each factor $t$ in
the time-Taylor expansion
\eqref{shorttimeexpansion} are either $e^{-i\twop \cdot \z }$
or $e^{-i\twoq \cdot \z } $. In the limit $\vert \y \vert \to \infty
$, $t \to 0$ we can make such  terms finite
 by shifting simultaneously $\twop \cdot \y $ and $\twoq
\cdot \y$ by $\ln t$.

The similarity ansatz as explained above is however not applicable to
the case of more than two initial modes. Instead, we have to reduce
the multimode initial condition to various two-mode problems  which 
can be handled in the usual way.  Let us illustrate this by looking at a
simple three-mode initial condition
\begin{equation}
\label{e:three_modes}
\Psi _0 (\x ) = h_1 e^{i\twop \cdot \x } + h_2 e^{i\twoq \cdot \x } +
+ h_3 e^{i\twor \cdot \x } + \mathrm{c.c.}\;,
\end{equation}
in which the vectors $\twop$, $\twoq$ and $\twor$ are listed in angular
counterclockwise order.
As in Section~\ref{s:pshydr}, to avoid pathologies, we assume
that the vectors $\twop$, $\twoq $, $\twor $ are not parallel and not
of the same length. In the Taylor expansion
\eqref{shorttimeexpansion} each factor $t$ will now be accompanied by a
factor $e^{-i\twop \cdot \z }$, $e^{-i\twoq \cdot \z }$ or
$e^{-i\twor \cdot \z }$.  Note that, in the limit $|\y| \to \infty $, $t
\to 0$ we cannot simultaneously 
make the terms $te^{\twop \cdot \y } $, $te^{\twoq
\cdot \y } $ and $te^{\twor \cdot \y } $  remain finite.
Indeed, we cannot translate $\y$ in such a way that all three
scalar products $\twop
\cdot \y $, $\twoq \cdot \y$ and $\twor \cdot \y$ are
shifted by $\ln t$.  
If $\pf _2 / \pf _1 \leq y_2 / y_1 \leq \qf
_2 / \qf _1 $ the factors $te^{\twop \cdot \y } $ and
$te^{\twoq \cdot \y } $ will dominate, while if
$\qf _2 / \qf _1 \leq y_2 / y_1 \leq \rf _2 / \rf _1 $ the factors
$te^{\twoq \cdot \y } $ and  $te^{\twor \cdot \y } $ will dominate.
In each case the three-mode initial condition \eqref{e:three_modes}
is reduced to a two-mode problem  involving either $\twop $ and
$\twoq $ or $\twoq $~and~$\twor $.

Let us now turn to the general multimode case with an
initial stream function of the form
\begin{equation} \label{e:polynomial}
 \Psi _0 (z_1 , z_2 ) = \sum_{(k_1 , k_2) \in \mathrm{supp} \, \hat{F} ^{(0)}
} \hat{F} ^{(0)} (k_1 , k_2 ) e^{-ik_1 z_1 } e^{-ik_2 z_2 } \;.
\end{equation}
Here, we assume Hermitian symmetry\footnote{In fact, this 
condition is not essential. We could as well consider more general sets 
$\mathrm{supp} \, \hat{F} ^{(0)}$.}  $\hat{F} ^{(0) \, \ast } (k_1 , k_2 ) = 
\hat{F} ^{(0)} (-k_1 , -k_2 )$ and we take the sum  over all wavevectors 
for which the Fourier coefficients $\hat{F} ^{(0)} (k_1 ,k_2)$ do not
vanish, called the \textit{support} of $\hat{F} ^{(0)} $ and denoted
$\mathrm{supp} \, \hat{F} ^{(0)}$.
The fact that $\mathrm{supp} \, \hat{F} ^{(0)} $ is a  finite set
plays a crucial part in our analysis. Let us suppose for simplicity
that all initial modes have different lengths, that is $\vert
(k_1^{\prime } , k_2^{\prime } ) \vert \neq \vert (k_1^{\prime \prime
} , k_2^{\prime \prime } ) \vert $ for all pairs $(k_1^{\prime } ,
k_2^{\prime } )$ and $(k_1^{\prime \prime } , k_2^{\prime \prime } ) \in
\mathrm{supp} \, \hat{F} ^{(0)}$.

As we have seen before, it is necessary to distinguish between
different directions in the $\y $-space when taking the limits $\vert
\y \vert \to \infty $, $t \to 0$.  Therefore we  let $\vert y_1
\vert , \vert y_2 \vert \to \infty $ while keeping  the ratio $y_2
/ y_1$  fixed. By Hermitian symmetry, 
it is enough to consider only the case $y_1 \to + \infty $.
\begin{figure}[t]
  \iffigs  
 \centerline{% 
 \includegraphics[scale=0.4]{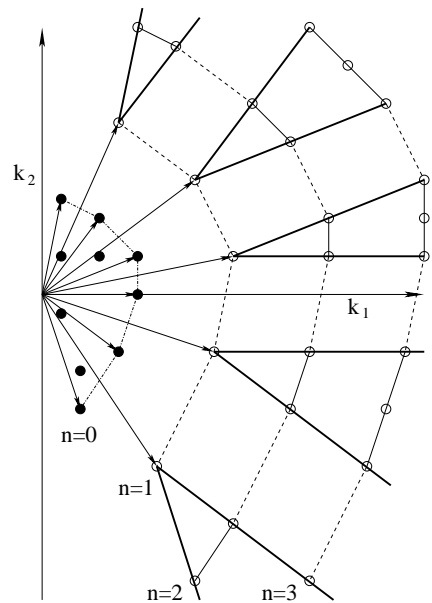}%
 }
 \else\drawing 65 10 {Multi mode picture}
 \fi  
 \caption{\label{f:multimode} Construction of relevant Fourier modes
   in various angular sectors.  Black circles: initial modes;
   dash-dotted lines: boundary of the Newton polytope of the initial
   modes; white circles: higher-order modes associated to $n$th
   generation. Thick lines show the edges of the various angular
   sectors.  }
\end{figure}
Assuming as in Section~\ref{s:pshydr} that 
\begin{equation}
\label{e:Taylor}
\Psi (\z ,t) = \sum_{n=0}^{\infty } \Psi _n (\z ) \, t^n\;,
\end{equation}
and denoting by $\hat{F} ^{(n)} (\k ) $ the Fourier coefficients of
$\Psi _n $ we obtain easily from the Euler equation the following
recursion relations for the Fourier coefficients of the \hbox{$(n+1)$th}
``generation'':
\begin{equation}
\label{e:Taylor_in_Fourier}
\begin{split}
& \hat{F} ^{(n+1)} (k_1 , k_2 ) = - \frac{1}{n+1} \frac{1}{\vert k \vert ^2 }
\sum_{m+p=n} \sum_{\k ^{\prime } + \k ^{\prime \prime } = \k } \\ & (\k
^{\prime } \wedge \k^{\prime \prime }) \vert \k ^{\prime \prime } \vert ^2
\hat{F} ^{(m)} (k_1^{\prime } , k_2^{\prime } ) \hat{F} ^{(p)} (k_1^{\prime
\prime } , k_2^{\prime \prime } ) \;,
\end{split}
\end{equation}
which allows us to compute $\hat{F} ^{(n+1)} (\k ) $ in terms of the
previous generations $\hat{F} ^{(m)} (\k ) $ and $\hat{F} ^{(p)} (\k )
$ with $m,p \geq 0$ and $m+p=n$.  From \eqref{e:Taylor_in_Fourier}
follows immediately that $\hat{F} ^{(n)} $ has finitely many
non-vanishing modes

We  now identify the modes in the $n$th generation which give the
leading-order contributions to $\Psi _n (\z )$ for a fixed ratio $y_2
/y_1 $. For this  we use the notion of {\it Newton polytope} of
$\mathrm{supp} \, \hat{F} ^{(0)} $.  It is defined as the convex hull
(in the usual sense) of the set $\mathrm{supp} \, \hat{F} ^{(0)} $, as
for example represented (taking into account the Hermitian symmetry)
on Fig.~\ref{f:multimode} for a typical initial condition. We shall
call {\it relevant} those initial modes lying on the boundary of the Newton
polytope of $\mathrm{supp} \, \hat{F} ^{(0)} $ (black circles
indicated by arrows on Fig.~\ref{f:multimode}). The relevant modes
divide the $\k $-space into angular sectors, analogously to the
three-mode case presented above.

Let now $\k ^{\prime } $ and $\k ^{\prime \prime } $ be two relevant
modes defining an angular sector such that $\k ^{\prime } \wedge \k
^{\prime \prime } > 0 $. The vectors $\k ^{\prime } $ and $\k ^{\prime
\prime } $ define an angular sector in the $\y $-space such
that $k_2^{\prime } /k_1^{\prime } \leq y_2 / y_1 \leq k_2^{\prime
\prime } / k_1^{\prime \prime } $. For these directions the leading-order 
terms in $\Psi _n (\z ) $ are proportional to $\ue ^{(n^{\prime }
\k ^{\prime } + n^{\prime \prime } \k ^{\prime \prime } )\cdot \y }$
with $n^{\prime } ,n^{\prime \prime } \geq 1$ and $n^{\prime } +
n^{\prime \prime } = n + 1 $; the terms corresponding to other Fourier
modes are subdominant. Clearly, we would have obtained the same
dominant terms if we had started from just the two initial modes $\k
^{\prime } $ and $\k ^{\prime \prime } $. We can now apply the
similarity ansatz, shifting $\k ^{\prime } \cdot \y $ and $\k ^{\prime
\prime } \cdot \y $ by $\ln t$. Let us remark that while it is possible
to eliminate the time variable by a global similarity ansatz for
two-mode asymptotics, it is in general impossible to do so for more
than two modes.

In the exceptional cases where two or more relevant modes have the
same length one must take into account the fact that the Fourier
coefficient of their sum vanishes.  The description of the
leading-order 
contributions in the $n$th generation becomes slightly more
involved than in the generic case, but the leading-order behavior in a
fixed direction $y_2 / y_1 $ is still dominated by two-mode
asymptotics.

Summarizing, we have shown that the study of the short-time asymptotics of
Euler flows with multimode initial conditions can be reduced to the analysis
of various two-mode asymptotics in angular sectors defined by a suitable set
of relevant initial modes, namely those on the boundary of the Newton polytope
of the initial modes.


\begin{thebibliography}{50}
\bibitem{poem}
And it bubbles and seethes, and it hisses and roars,\\
As when fire is with water commix'd and contending,\\
And the spray of its wrath to the welkin up-soars,\\
And flood upon flood hurries on, never-ending.\\
(translation by Edward, Lord Lytton.)\\
\end{thebibliography}
\end{document}